\documentclass[aps,pra,twocolumn,superscriptaddress]{revtex4-1}%

\usepackage{bm}
\usepackage{chapterbib}

\usepackage{amsmath,amssymb}
\usepackage{amsfonts}
\usepackage{layouts}
\usepackage{mathtools}
\usepackage{dsfont}
\usepackage{graphicx}
\usepackage{float}
\usepackage{subfigure} 
\usepackage{verbatim}
\usepackage{amsfonts}
\usepackage{bbold}
\usepackage{dcolumn}
\usepackage{bm}
\usepackage{color}
\usepackage[dvipsnames]{xcolor}
\usepackage{listings}
\definecolor{blueprl}{RGB}{46,48,146}

\usepackage{graphics}
\usepackage{amsmath}
\usepackage{amssymb}
\usepackage{longtable}
\usepackage{epsfig}
\usepackage{latexsym}
\usepackage{bbm}
\usepackage{psfrag}%
\usepackage{mathrsfs}
\newcommand{\bra}[1]{\mbox{$\langle #1 |$}}
\newcommand{\ket}[1]{\mbox{$| #1 \rangle$}}
\def\ketbra#1#2{{\vert#1\rangle\!\langle#2\vert}}

\newcommand{\xdownarrow}[1]{%
  {\left\downarrow\vbox to #1{}\right.\kern-\nulldelimiterspace}
}

\makeatletter
\newcommand*{\balancecolsandclearpage}{%
  \close@column@grid
  \clearpage
  \twocolumngrid
}
\makeatother

\usepackage{gensymb}
\usepackage[breaklinks=true]{hyperref}
\hypersetup{
colorlinks   = true, 
urlcolor     = black, 
linkcolor    = black, 
citecolor    = black 
}
\usepackage{comment}
\usepackage[normalem]{ulem}
\usepackage{xcolor}
\usepackage{graphicx}
\usepackage{cleveref}
\crefname{equation}{equation}{equations}
\Crefname{equation}{Equation}{Equations}
\crefname{figure}{Fig.}{Figs.}
\Crefname{figure}{Figure}{Figures}
\crefname{figure}{Fig.}{Figs.}
\Crefname{figure}{Figure}{Figures}
\crefname{section}{}{}
\Crefname{section}{}{}
\crefname{appendix}{Appendix}{Appendices}
\Crefname{appendix}{Appendix}{Appendices}
\crefname{table}{Table}{Tables}
\Crefname{table}{Table}{Tables}

\newcommand\NLAto{\stackrel{\mathclap{\footnotesize	\mbox{NLA}}}{\longmapsto}}

\newcommand\lossto{\stackrel{\mathclap{\footnotesize	\mbox{loss}}}{\longmapsto}}

\newcommand{\bblack}[1]{\textcolor{black}{#1}}

\newcommand{\rred}[1]{\textcolor{black}{#1}}

\newcommand{\Tr}{\text{Tr}}
\newcommand{\tr}{\text{tr}}

\newcommand{\Id}{\mathbb{1}}

\makeatletter
\def\@bibdataout@aps{%
 \immediate\write\@bibdataout{%
  @CONTROL{%
   apsrev41Control,author="08",editor="1",pages="0",title="0",year="1",eprint="1"%
  }%
 }%
 \if@filesw
  \immediate\write\@auxout{\string\citation{apsrev41Control}}%
 \fi
}%
\makeatother 

\usepackage[hyphenbreaks]{breakurl}
\usepackage{enumitem}

\begin{document}

\title{Achieving the ultimate end-to-end rates of lossy quantum communication networks
}

\author{Matthew S. Winnel}\email{mattwinnel@gmail.com}
\affiliation{Centre for Quantum Computation and Communication Technology, School of Mathematics and Physics, University of Queensland, St Lucia, Queensland 4072, Australia}
\author{Joshua J. Guanzon} 
\affiliation{Centre for Quantum Computation and Communication Technology, School of Mathematics and Physics, University of Queensland, St Lucia, Queensland 4072, Australia}
\author{Nedasadat Hosseinidehaj} 
\affiliation{Centre for Quantum Computation and Communication Technology, School of Mathematics and Physics, University of Queensland, St Lucia, Queensland 4072, Australia}
\author{Timothy C. Ralph}
\affiliation{Centre for Quantum Computation and Communication Technology, School of Mathematics and Physics, University of Queensland, St Lucia, Queensland 4072, Australia}

\date{\today}

\begin{abstract}
The field of quantum communications promises the faithful distribution of quantum information, quantum entanglement, and absolutely secret keys, however, the highest rates of these tasks are fundamentally limited by the transmission distance between quantum repeaters. \bblack{The ultimate end-to-end rates of quantum communication networks are known to be achievable by an optimal entanglement distillation protocol followed by teleportation. In this work, we give a practical design for this achievability. Our ultimate design is an iterative approach, where each purification step operates on shared entangled states and detects loss errors at the highest rates allowed by physics. As a simpler design, we show that the first round of iteration can purify completely at high rates. We propose an experimental implementation using linear optics and photon-number measurements} which is robust to inefficient operations and measurements, showcasing its near-term potential for real-world practical applications.
\end{abstract}

\maketitle

The great challenge for quantum communication~\cite{Gisin_2007} is how to overcome loss~\cite{book}, the dominant source of noise through free space and telecom fibers. Many applications~\cite{PhysRevLett.120.080501,PhysRevLett.121.043604,PhysRevA.97.032329,7562346,DANOS200773}, including quantum key distribution (QKD)~\cite{Pirandola_2020,Xu_2020} (i.e., the task of sharing a secret random key between two distant parties), suffer from an \bblack{exponential rate-distance scaling~\cite{PhysRevLett.102.050503,takeoka2014}}. Determining the most efficient protocols for distributing quantum information, entanglement, and secure keys is of vital importance to realise the full capability of the quantum internet~\cite{kimble2008}.

\bblack{It is known that the reverse coherent information (RCI)~\cite{PhysRevLett.102.210501} is an achievable rate for entanglement distillation by an \rred{implicit optimal protocol based on \textit{one-way} classical communication}. For the bosonic pure-loss channel, this rate is $R = -\log_2{(1-\eta)}$~\cite{PhysRevLett.102.050503}, where $\eta \in [0,1]$ is the channel transmissivity. This is an achievable rate for entanglement distillation, $E_{\text{D}}$, over the lossy channel and is also an achievable rate for secret key distribution, $K$, since an ebit is a specific form of secret key bit. To summarise, we have $K \geq E_{\text{D}} \geq R = -\log_2{(1-\eta)}$. Ref.~\cite{Pirandola_2017} proved the upper bound, the so called Pirandola–Laurenza–Ottaviani–Banchi (PLOB) bound, that is, $K \leq  -\log_2{(1-\eta)}$. This, together with the lower bound, $R$, from Ref.~\cite{PhysRevLett.102.050503}, establishes $K = E_{\text{D}} = R = -\log_2{(1-\eta)} = C$, the two-way assisted entanglement distribution capacity and secret key distribution capacity of the pure-loss channel.}


Likewise, there are fundamental limits to the highest end-to-end rates of arbitrary quantum communication networks~\cite{Pirandola_2019}, where untrusted quantum repeaters divide the total distances into shorter quantum channels (links). Quantum repeaters are strictly required to beat the PLOB bound~\cite{7010905,Muralidharan2016}. For a linear repeater chain, it is optimal to place repeaters equidistantly, then the ultimate end-to-end rate is given by $-\log_2{(1-\eta)}$~\cite{Pirandola_2019}, where $\eta$ now refers to the transmissivity of each link. For a multiband network, consisting of $m$ generally-entangled channels in parallel, the rate is additive, $-m\log_2{(1-\eta)}$~\cite{Pirandola_2019}. \bblack{These ultimate repeater bounds are achievable by using an optimal entanglement distillation protocol followed by quantum teleportation (entanglement swapping), while ideal quantum memories are most-likely required to achieve the highest rates.}  

\bblack{The goal of this paper is to give a physical realisation for achieving these ultimate rates, which could pave the way for experimental implementations. While the highest achievable secret key rate for point-to-point CV QKD saturates the PLOB bound~\cite{Pirandola_2020}, it does not provide a physical design for entanglement distillation.} Furthermore, \bblack{it is impossible to distil Gaussian entanglement using Gaussian operations only~\cite{PhysRevA.66.032316,PhysRevLett.89.137903}} so quantum repeaters must use non-Gaussian elements~\cite{PhysRevA.90.062316}.

Protocols based on infinite-dimensional systems are required to saturate the ultimate limits. However, the majority of quantum-information-processing tasks and techniques are for discrete-variable (DV) systems~\cite{10.5555/1972505} where the quantum information is finite dimensional. In contrast, for continuous-variable (CV) systems~\cite{Braunstein_2005,yonezawa2008continuousvariable,book,Weedbrook_2012,Serafini2017QuantumCV}, the quantum information is infinite dimensional and encoded in the quadrature amplitudes, and in principle offer easier state manipulation~\cite{Braunstein_2005} and compatibility with existing optical telecom infrastructure~\cite{Kumar_2015}. \bblack{Previous practical quantum repeater designs are unable to distil entanglement at the ultimate rates~\cite{dias2017,Furrer2018,PhysRevResearch.2.013310,Ghalaii_2020,Dias_2020,winnel2021overcoming}.}

Quantum repeaters have previously been categorised into three generations depending on how they combat loss and other sources of noise~\cite{7010905,Muralidharan2016}. With respect to CV systems, the first two generations~\cite{ralph2009nondeterministic,PhysRevA.102.063715,guanzon2021ideal,fiurasek2021teleportationbased,PhysRevA.86.012327, PhysRevA.89.023846, PhysRevA.91.062305} remove loss via teleportation-based techniques, for instance, entanglement swapping and/or noiseless linear amplification~\cite{ralph2009nondeterministic,PhysRevA.102.063715,guanzon2021ideal,fiurasek2021teleportationbased,PhysRevA.86.012327, PhysRevA.89.023846, PhysRevA.91.062305}. These techniques fail to achieve the ultimate limits under pure loss since the output state is not pure and the success probability is zero for high-energy input states. For instance, the schemes based on noiseless linear amplification~\cite{Ghalaii_2020,Dias_2020,winnel2021overcoming} have the same rate-distance scaling as the ultimate bounds but do not saturate them. A simple explanation is given in Supplementary Note 1, also see Ref.~\cite{Pandey_2013}.

The third generation of quantum repeaters~\cite{surfact_code_2010} uses quantum error correction~\cite{gottesman2009introduction} and is a purely one-way communication scheme. It promises high rates since it does not require back-and-forth classical signalling, however, here the ultimate rates are bounded by the unassisted quantum capacity of each link~\cite{Pirandola_2017}, $  \log_2{( \frac{\eta}{1-\eta} )} <  C$. This means the third-generation rate is zero if $\eta \leq 0.5$, which translates to a maximum link distance of about 15 km for optical fiber with a loss rate of 0.2 dB/km. In contrast, the two-way assisted capacity of pure loss allows a nonzero achievable rate at all distances. It is interesting to note that the family of GKP codes~\cite{GKP2001} achieves the unassisted capacity of general Gaussian thermal-loss channels with added thermal noise, where pure loss is the zero-temperature case, up to at most a constant gap~\cite{GKP2019}. \bblack{Likewise, our main result is to give a practical protocol which achieves the two-way assisted capacity of the pure-loss channel.}

In summary, all three generations of quantum repeaters are unable to operate at rates which saturate the ultimate limits of quantum communications. Motivated by this reality, we introduce an iterative protocol to \bblack{purify completely} from pure loss and achieve the capacity of the channel. Our schemes are inspired by Refs.~\cite{Bennett_1996,PhysRevLett.84.4002}. \bblack{The idea is that neighbouring nodes locally perform photon-number measurements on copies of shared CV entanglement across the lossy channel, followed by \rred{\textit{two-way}} classical communication to compare photon-number outcomes.} We show that the highest rates of our purification scheme, requiring \rred{two-way} classical communication, achieve the fundamental limits of quantum communications for pure-loss channels. In contrast to quantum error correction, we describe purification as a quantum-error-detection scheme against loss. \rred{We consider a much-simpler design with good rates requiring only \textit{one-way} classical communication and no iteration.}

In this work, the required measurements are quantum non-demolition measurements (QND) of total photon number of multiple modes and can be implemented experimentally using \bblack{linear optics and photon-number measurements}. This implementation is naturally robust against inefficiencies of the detectors and gates. Alternatively, these QND measurements can be implemented using high finesse cavities and cross–Kerr nonlinearities~\cite{PhysRevLett.84.4002}.

\section*{Results}

First, we introduce our iterative protocol for the complete purification of high-dimensional entanglement, saturating the two-way assisted capacity of the bosonic pure-loss channel. Then, we \bblack{show that our} protocol without iteration \bblack{(i.e. single-shot)} still gives high rates which fall short of the ultimate limits by at worst a factor of $0.24$. Finally, we explain how to implement our protocol using \bblack{linear-optics and photon-number measurements}.

\bigskip

\textbf{Iterative purification.} Alice and Bob share multiple copies of a state which is entangled in photon number, such that in a lossless situation they will always measure the same number of photons. Our purification technique, in a pure loss situation, is for Alice and Bob to each locally count the total number of photons contained in multiple copies of the shared entangled states, and then compare the results. If they \rred{locally} find a different total number of photons, this means photons were lost. Alice and Bob then iteratively perform total photon-number measurements over smaller subsets of states until their outcomes are the same, and hence distil pure entanglement. We prove that the highest average rate of the protocol achieves the capacity of the pure-loss channel.

\begin{figure}
\includegraphics[width=0.9\linewidth]{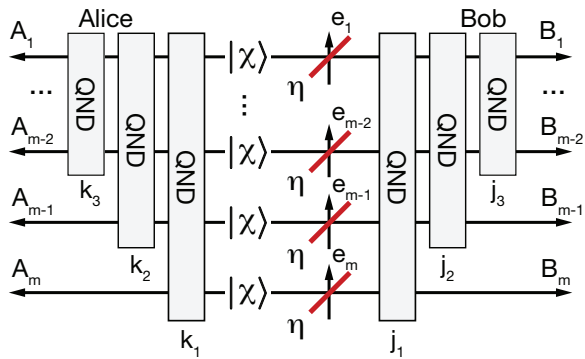}
\caption{\textbf{Our iterative protocol for the complete purification of Gaussian continuous-variable quantum states}. Alice shares $m$ two-mode-squeezed-vacuum states with Bob across independent pure-loss channels with transmissivity $\eta$. \bblack{Iterative QND measurements of total photon number at Alice's and Bob's sides followed by classical communication herald pure states whenever $k_n=j_n$, which means round $n$ is successful and the protocol is complete. The highest average rate of quantum communication achieves the capacity (the PLOB bound~\cite{Pirandola_2017}).} Alice's measurements encode the quantum information onto the rails, and Bob's measurements purify the entanglement and decode the quantum information.}
\label{fig:QND}
\end{figure}

We now consider our protocol in detail. The protocol is shown in~\cref{fig:QND}. Consider two neighbouring nodes in a network, Alice and Bob, separated by a repeaterless link. \bblack{Round one of our iterative protocol is identical to entanglement purification of} Gaussian CV quantum states from Ref.~\cite{PhysRevLett.84.4002}, however, our protocol includes an iterative procedure.  Alice prepares $m$ copies of a pure two-mode squeezed vacuum (TMSV) state, $\ket{\chi}=\sqrt{1-\chi^2} \sum_{n=0}^{\infty} \chi^n \ket{n}\ket{n}$ in the Fock photon-number basis, with squeezing parameter $\chi \in [0,1]$. \bblack{The unique entanglement measure, $E$, for a bipartite pure state, $\ket{\phi}$, is given by the von Neumann entropy, $S$, of the reduced state, i.e., $E =-\tr{(\rho_A \log_2{\rho_A})}$, where $\rho_A=\tr_B{\ketbra{\phi}{\phi}}$. This means Alice initially prepares $m E_\chi$ ebits of entanglement, where $E_\chi = G[(\lambda_k{-}1)/2]$, where $G\rred{(x)}=(x{+}1)\log_2(x{+}1){-}x\log_2(x)$, $\lambda_k=2\bar{n}{+}1$, $\bar{n} = \sinh(r)^2$, and $r = \tanh^{-1}{\chi}$.}

Alice shares the second mode of each of the $m$ pairs with Bob across the link.  The error channel we consider is bosonic pure loss, modelled by mixing the data rails with vacuum modes of the environment, or a potential eavesdropper (Eve), on a beamsplitter with transmissivity $\eta$.

Alice encodes the quantum information into a quantum-error-detecting code so that Bob can detect errors on his side. To do this, she performs a QND measurement of total photon number on the $m$ modes and obtains outcome $k_1$ \bblack{(where the subscript $1$ refers to the round one of iteration)}, and shares this information with Bob via classical communication. Alice's measurement projects the system before the channel onto a maximally-entangled state~\cite{PhysRevLett.84.4002}
\begin{align}
    \ket{\phi_{k_1,m}}_{AB} &= (1-\chi^2)^\frac{m}{2} \chi^{k_1} \stackrel{n_{1}+n_{2}+\cdots +n_{m}=k_1}{\mathrel{\mathop{\sum }\limits_{n_{1},n_{2},\cdots ,n_{m}}}} \nonumber \\
    &\;\;  \ket{ n_{1},n_{2},\cdots ,n_{m}}_{A} \ket{ n_{1},n_{2},\cdots ,n_{m}}_{B} \\
    &= (1{-}\chi^2)^\frac{m}{2} \chi^{k_1}    \sum_{\mu=0}^{d_{k_1,m}{-}1}   \ket{ \mu_{k_1,m}}_A \ket{ \mu_{k_1,m}}_{B},
\end{align} where $ \ket{\mu_{k_1,m}} = \ket{n_1^{(\mu)},n_2^{(\mu)},\cdots,n_m^{(\mu)}}$ can be viewed as orthogonal basis states which form a quantum-error-detecting code, each composed of $\sum_{i=1}^m n_i^{(\mu)}=k_1$ photons. We discuss the code in detail later. The pure maximally-entangled state $\ket{\phi_{k_1,m}}_{AB}$ has entanglement $E_{k_1,m}=\log_2{d_{k_1,m}}$ ebits with dimension $d_{k_1,m} = {k_1+m-1 \choose k_1}$, and the probability of Alice's measurement outcome, $k_1$, is $P_{k_1,m}^{\text{Alice}} = (1-\chi^2)^m \chi^{2 k_1} d_{k_1,m}$. The dimension, $d_{k_1,m}$, is the total number of ways $k_1$ identical photons can be arranged among the $m$ distinct rails.

Bob then performs a QND measurement of total photon number across the $m$ rails on his side to detect loss errors. If Bob obtains the outcome $j_1$ photons, and knows that Alice sent $k_1$ photons, then both Alice and Bob know that $k_1-j_1$ photons were lost. Bob's QND measurement, with success probability $P_{k_1,j_1}^\text{Bob}=(1{-}\eta)^{k_1{-}j_1} \eta^{j_1} {k_1 \choose j_1}$, heralds a renormalised mixed state shared between Alice and Bob which does not depend on loss for any outcome $j_1$. That is, together Alice's and Bob's QND measurements remove all dependence on loss and has exchanged success probability for entanglement, while they learn that $k_1-j_1$ photons were lost to the environment. All outcomes besides zero at Alice and Bob herald useful entanglement. If $j_1\neq k_1$, they must do further rounds of purification (iteration) since the output state is not pure. If $j_1=k_1$, then the output state is strictly pure and purification is complete in a single round. For a simpler protocol, Alice and Bob may post select on outcomes $j_1=k_1$ without further iteration. We show later in the paper that this single-shot protocol still gives excellent rates.

Explicitly, the global output state heralded by outcomes $k_1$ and $j_1$ is
\begin{align}
&\ket{\phi_{k_1,j_1,m}}_{ ABe }  =  (1-\chi^2)^{\frac{m}{2}} \chi^{k_1} (1-\eta)^\frac{k_1-j_1}{2} \eta^\frac{j_1}{2} \nonumber \\
&\;\;\;\;\;\;\; \stackrel{n_{1}+n_{2}+\cdots +n_{m}=k_1}{\mathrel{\mathop{\sum }\limits_{n_{1},n_{2},\cdots ,n_{m}}}} \;  \nonumber \stackrel{l_1+l_2+\cdots+l_m=k_1-j_1,\;\; l_i\leq n_i \forall i}{\mathrel{\mathop{\sum }\limits_{l_{1},l_{2},\cdots ,l_{m}}}} \;  \nonumber \\ &\;\;\;\;\; \sqrt{{n_1 \choose l_1}  {{n_2 \choose l_2}} {{n_3 \choose l_3}} \cdots {{n_m \choose l_m}}} \;   \ket{ n_{1},n_{2},\cdots ,n_{m}} _{A} \nonumber \\
&\;\;\;\;\;\;\;\;\;\;  \otimes \ket{ n_{1}-l_{1},n_{2}-l_{2},\cdots ,n_{m}-l_{m}}_{
B } \nonumber \\
&\;\;\;\;\;\;\;\;\;\;\;\;\; \otimes \ket{ l_{1},l_{2},\cdots ,l_{m}}_{e },\label{eq:phi_jk}
\end{align}
where $A,B,e$ refer to the $m$-rail quantum systems owned by Alice, Bob, and the environment, respectively, as shown in~\cref{fig:QND}. The full derivation of this state is in Supplementary Note 2. The factor $(1-\eta)^\frac{k_1-j_1}{2} \eta^\frac{j_1}{2}$ is outside the sum, thus, we have the remarkable result that the renormalised output state shared between Alice and Bob does not depend on $\eta$. Therefore, the entanglement shared between Alice and Bob also has no dependence on $\eta$, which has been exchanged for probabilities.

\bblack{Additional rounds of purification can purify more entanglement after the initial round. One approach is for Alice and Bob to locally perform QND measurements as in round one but on $m-n+1$ rails, and obtain outcomes $k_n,j_n$, where $n$ refers to the round number. \rred{At round $n$, there is no entanglement shared between Alice and Bob on the last $n-1$ rails and the photon number of each of these rails is completely known. At round $n$, these last $n-1$ rails can be discarded while the $m-n+1$ rails should be kept.}
}

\bigskip

\textbf{Exact rate of iterative entanglement purification for finite numbers of rails.}
The rate of our purification protocol (in ebits per use) is maximised if Alice performs her first measurement offline (i.e., setting $P_{k_1,m}^{\text{Alice}}=1$) where she obtains outcome $k_1$. For finite $m$, there is a finite $k_1$ which optimises the rate. \rred{However, for large squeezing $\chi\to1$ outcomes $k_1$ are dominated by $k_1\to\infty$ with unity probability. Therefore, the large squeezing limit $\chi\to\infty$ without $k_1$ pre-selection, is equivalent to $k_1\to\infty$ with offline $k_1$ pre-selection.}

\rred{Taking Alice’s first measurement to be done with result $k_1$ offline (which can be chosen in advance to optimise the rate or, for example, the practicality of the protocol)}, the rate for finite $m$ is
\begin{align}
    E_{k_1,m}(\eta) &= \frac{1}{m} \sum_{n=1}^{m-1} \sum_{j_1,k_2,j_2,\dots,k_n,j_n} P E, \label{eq:exact_rate_finite_m}
\end{align}
where the sum is constrained by
\begin{align}
    &k_1 \geq k_2 \geq k_3 \geq \dots \geq k_n,\\
    &j_1 \geq j_2 \geq j_3 \geq \dots \geq j_n,\\
    &k_s-k_{s+1} \geq j_s-j_{s+1} \; \forall s ,\\
    &k_s > j_s \; \forall s \neq n, \\
    &k_n=j_n,
\end{align}
where the probability of success for a particular combination of outcomes, $j_1,k_2,j_2,\dots,k_n,j_n$, for a given $k_1$ and $m$ is
\begin{align}
      P &= (1-\eta)^{k_1-j_1} \eta^{j_1}   \frac{ {{k_n+m-n} \choose k_n} {k_n \choose j_n} }{ {k_1+m-1 \choose k_1} }  \left[ \prod_{s=1}^{n-1} {{k_s-k_{s+1}} \choose {j_s-j_{s+1}}} \right] ,\label{eq:probability}
\end{align}
where a maximally-entangled state is generated with entanglement
\begin{align}
      E &= \log_2{\left[ { {{k_n+m-n} \choose k_n} } \right]}.\label{eq:entanglement}
\end{align}
The rate $E_{k_1,m}(\eta)$ for finite $m$ can be optimised over Alice's initial outcome $k_1$ prepared offline and the number of rails $m$ as a function of $\eta$. We numerically compute the rate in Supplementary Note 3 for small $k_1$ and $m$. \rred{We show next that the highest rate of our iterative protocol achieves the capacity, $C$, for $m\to\infty$ and $k_1\to\infty$ (i.e., $\chi\to1$). We show this without having to compute~\cref{eq:exact_rate_finite_m} directly which would be arduous. }


\bigskip

\textbf{Optimality of our protocol.} The RCI~\cite{PhysRevLett.102.050503,PhysRevLett.102.210501}, $R$, gives an achievable lower bound on the distillable entanglement, $E_{\text{D}}$, and on the optimal secret key rate. The RCI is defined in the Methods section. We will show that our protocol is optimal for entanglement distillation as $m\to\infty$ and $\chi\to1$ and that no entanglement is lost \rred{between rounds}. \rred{We use the RCI as a benchmark to test the quality of our distillation procedure. The optimal distillation protocol implicit by the RCI is not required here since our scheme gives the same performance as the implicit optimal protocol round after round for large $m$.}

We require that the \rred{weighted} average von Neumann entropy of the \rred{reduced} pure states heralded after round one, $S_1$,  plus the average RCI of the failure states heralded after round one, $F_1$, equals the RCI of the state before round one. Note that the RCI equals the von Neumann entropy for pure states. We have
\begin{align}
    S_1+F_1 &= \frac{1}{m} \sum_{k_1=0}^\infty \sum_{j_1=0}^{k_1} P_{k_1,j_1,m} R_{k_1,j_1,m},\label{eq:sum}
\end{align}
in units of ebits per channel use, \rred{where
\begin{align}
    S_1 &= \frac{1}{m} \sum_{k_1=0}^\infty  P_{k_1,j_1=k_1,m} R_{k_1,j_1=k_1,m},\\
    F_1 &= \frac{1}{m} \sum_{k_1=0}^\infty \sum_{j_1=0}^{k_1-1} P_{k_1,j_1,m} R_{k_1,j_1,m},
\end{align}
}where $P_{k_1,j_1,m}  {=} P_{k_1,m}^{\text{Alice}}  P_{k_1,j_1}^\text{Bob}$, where $P_{k_1,m}^{\text{Alice}}=(1{-}\chi^2)^m \chi^{2k_1} d_{k_1,m}$ and $P_{k_1,j_1}^\text{Bob}=(1{-}\eta)^{k_1{-}j_1} \eta^{j_1} {k_1 \choose j_1}$. 
\rred{When the first round succeeds, the entanglement of the renormalised maximally-entangled pure-state shared between Alice and Bob heralded by outcomes $k_1=j_1$ is given by the von Neumann entropy of the reduced state:
\begin{align}
     E_{k_1=j_1,m} &= R_{k_1,j_1=k_1,m} =  \log_2{\left[ { {k_1+m-1 \choose k_1} } \right]},\label{eq:E_kj}
\end{align}
}which does not depend on the transmissivity, $\eta$, nor the amount of two-mode squeezing, $\chi$. When the first round fails, the RCI of the renormalised mixed state shared between Alice and Bob heralded by each pair of outcomes $k_1$ and $j_1$ is
\begin{align}
    R_{k_1,j_1,m} =  \log_2{\left[ \frac{ {k_1+m-1 \choose k_1} }{ {k_1-j_1+m-1 \choose k_1-j_1} } \right]},\label{eq:E_kj}
\end{align}
which also does not depend on the transmissivity, $\eta$, nor the amount of two-mode squeezing, $\chi$. See Supplementary Note 2 for the derivation.

\rred{The amount of squeezing is scaled to infinity $\chi\to1$, such that Alice initially measures a large amount of photons $k_1\to\infty$ and $m/k_1\to0$. Furthermore, from the fact that $P_{k_1,j_1}^\text{Bob}$ is a binomial distribution, Bob will most likely measure $j_1 \approx \eta k_1$ photons. Using these conditions, we show in Supplementary Note 3 that~\cref{eq:sum} approaches}
\begin{align}
   \lim_{\chi\to1} (S_1+F_1) 
   &= - \frac{m-1}{m} \log_2{(1-\eta)} = \frac{m-1}{m} C,
\end{align}
\bblack{which ensures that round one of purification is optimal since there is no loss of rate after round one as $m\to\infty$, given that the average RCI at the end of the round equals the initial RCI of the protocol. Thus, there exists an optimal protocol to follow round one which can saturate the two-way assisted quantum capacity (the PLOB bound) using our protocol as an initial step.} The entanglement is \textit{optimally} exchanged for success probability.

\bblack{Similarly, we prove in Supplementary Note 3 that round $n$ is optimal since there is no loss of rate at round $n$ as $m\to\infty$, up to the same factor, $\frac{m-1}{m}$. This factor comes from Alice's and Bob's measurement of photon number.}

To quantify this loss of entanglement, consider the protocol before the channel. We'll see that for finite $m$ some entanglement is immediately lost after Alice's QND measurement. \rred{Ref.~\cite{PhysRevLett.84.4002} defined the entanglement ratio, denoted by $\Gamma_{1}$}, as the average entanglement heralded by Alice's QND measurement divided by the total initial entanglement $m E_\chi$, that is,
\begin{align}
    \Gamma_{1} &\rred{\equiv} \frac{\sum_{k_1=0}^\infty P_{k_1,m}^{\text{Alice}} E_{k_1,m}}{m E_\chi}.
\end{align}
In the limit of a large number of rails $\lim_{m\to\infty} \Gamma_1 = 1$ for all $0<\chi\leq1$, which means asymptotically Alice's QND measurement \rred{heralds no loss of} entanglement. However, for finite $m$ in the limit of large squeezing, $\lim_{\chi\to1} \Gamma_1 = (m-1)/m$. So, for finite number of rails $m$ some entanglement is lost since $1/2 < (m-1)/m <1$. This means we must take $m\to\infty$ to get the highest rates.

\rred{Similarly, to quantify the entanglement lost at round $n>1$, we define the entanglement ratio, $\Gamma_{k_{n-1}}$, as the weighted average entanglement heralded at round $n$ over all outcomes $k_n$ normalised by the weighted entanglement heralded by outcome $k_{n-1}$ at the previous round, $n-1$, that is,}
\begin{align}
   \rred{\Gamma_{k_{n-1}}} &\rred{\equiv} \rred{\frac{\sum_{k_n} P_{k_n} E_{k_n} }{ P_{k_{n-1}} E_{k_{n-1}}}}  =   \frac{ \sum_{k_n} d_{k_n} \log_2{d_{k_{n}}}}{d_{k_{n-1}}\log_2{d_{k_{n-1}}}},\label{eq:ratio_n}
\end{align}
\rred{where $d_{k_n} =  {{k_n+m-n} \choose k_n} $ and $d_{k_{n-1}} =  {{k_{n-1}+m-n+1} \choose k_{n-1}} $. $P_{k_n}$ ($P_{k_{n-1}}$) and $E_{k_{n}}$ ($E_{k_{n-1}}$) are defined in~\cref{eq:probability,eq:entanglement}, for a given $k_n$ ($k_{n-1}$), and of course we can take $j_i=k_i$ for all $i$ since here we consider no loss channel. Many of the factors cancel giving the simple expression in~\cref{eq:ratio_n}.} Curiously, for large numbers of rails the amount of entanglement lost at round $n$ is the same as for round one, i.e., $\lim_{m\to\infty} \rred{\Gamma_{k_{n-1}}} = (m-1)/m$ for \textit{all} $k_{n-1}$. \rred{This result ensures that ``encoding'' into $k_n$ photons is asymptotically optimal throughout the entire duration of our iterative procedure up to the factor $(m-1)/m$, which approaches unity for large $m$.}

\bigskip

\bblack{\textbf{Achieving the capacity.} The average \rred{rate in case of success} of pure entanglement distilled at round $n$ in ebits per use of the channel is
\begin{align}
    S_n = \frac{1}{m} \sum_{k_1,j_1,\dots,k_n,j_n}  P E\; \rred{\delta_{k_n,j_n}},
\end{align}
and the average \rred{rate in case of failure} of entanglement distilled at round $n$ in ebits per use of the channel is lower bounded by the average RCI
\begin{align}
    F_n = \frac{1}{m} \sum_{k_1,j_1,\dots,k_n,j_n} P R \; \rred{(1-\delta_{k_n,j_n})},
\end{align}
where $P$ is the probability from~\cref{eq:probability} and $R =  \log_2{\left[ \frac{ {k_n+m-n \choose k_n} }{ {k_n-j_n+m-n \choose k_n-j_n} } \right] }$ is the RCI of the heralded states. \rred{Note $\delta_{k_n,j_n}$ is the Kronecker delta function.} For the \rred{rate in case of success}, the RCI equals the von Neumann entropy since the states are pure, $R=E$. The entanglement \rred{in case of failure} at round $n$ will be purified at a later round.
}

\bblack{Since our protocol is optimal at round $n$ for $\chi\to 1$ and $m\to\infty$, we have the following expressions:
\begin{align}
    S_1 + F_1  &=   \frac{m-1}{m} C \\
    \lim_{m\to\infty} (S_{n} + F_{n} )&= \lim_{m\to\infty} \left( \frac{m-1}{m} F_{n-1}\right),\label{equalities}
\end{align}
for $2 \leq n \leq m$. We solve this system of equations by addition, and we find that the average \rred{rate in case of success} of \rred{purification using our iterative procedure for}
$m\to\infty$ and $\chi\to1$ is
\begin{align}
    E_\text{iteration} &= \lim_{m\to\infty}  \sum_{n=1}^m S_n\\ 
    &=  \lim_{m\to\infty} \frac{m-1}{m} C - \lim_{m\to\infty} \sum_{n=1}^{m-1}  \frac{F_n}{m}\\
    &= C.
\end{align}
We achieve the capacity (PLOB) in the limit of a large number of rails, where $\lim_{m\to\infty} \sum_{n=1}^{m-1}  \frac{F_n}{m} \to 0$ and $\lim_{m\to\infty} (m-1)/m\to1$. See Supplementary Note 3 for the proof.}

\bigskip

\textbf{Achievable rates of repeater networks.}

\begin{figure*}
\includegraphics[width=1\linewidth]{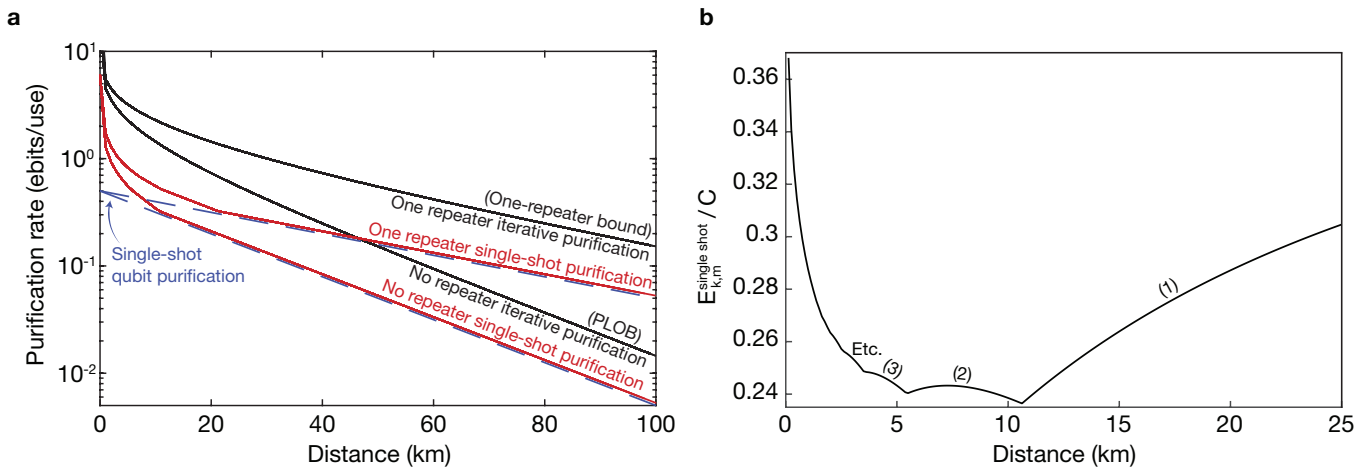}
\caption{\textbf{Optimal rates of entanglement purification.} \textbf{a} We plot the highest rates of iterative purification (black), single-shot purification (red), and single-shot purification for qubits (blue dashed) as a function of the total end-to-end distance, for optical fiber with a loss rate of 0.2 dB/km. Purification is used to distribute and purify entanglement between nodes and entanglement swapping connects the end users. To illustrate, we plot rates without a repeater and with one repeater. Equivalently, these rates are secret key rates since ebits are specific types of secret bits~\cite{secure_key_2005}. Our highest rates coincide with the ultimate limits of quantum communications (black), \bblack{shown here the repeaterless PLOB bound~\cite{Pirandola_2017} and the one-repeater bound~\cite{Pirandola_2019}.} At all distances the output states are strictly pure. The scaling improves for increased numbers of repeaters (not shown). \textbf{b} The ratio of the optimised rate of our single-shot purification protocol without a repeater with the repeaterless PLOB bound as a function of distance, showing how close it comes to saturating the bound (and therefore end-to-end quantum-repeater networks in general). The optimal number of photons, $k$, at each distance is shown in rounded brackets. At shorter distances codes with more photons are optimal. The three-rail encoding ($m=3$) is optimal at most distances, with sometimes a larger $m$ optimal at distances less than about 1.6 km. At large distances, the ratio $E_{k,m}^\text{single shot}/C$ approaches $\ln{3}/3 \sim 0.366$. As the distance goes to zero, $E_{k,m}^\text{single shot}/C=1/2$.}
\label{fig:optimal_rate}
\end{figure*}

Our protocol purifies completely at the PLOB rate. Assuming ideal quantum memories are available, after teleportation (entanglement swapping) we achieve the the ultimate end-to-end rates of quantum communication networks \rred{by adopting the routing methods of Ref.~\cite{Pirandola_2019}. That is, the results can be extended beyond chains to consider more complex topologies and routing protocols~\cite{Pirandola_2019}.} We describe details about entanglement swapping in Supplementary Note 4. We plot the highest rates of iterative purification as a function of total distance with no repeater and one repeater in~\cref{fig:optimal_rate}a (black lines) which coincide with the capacities. We also plot rates in~\cref{fig:optimal_rate}a for single-shot purification for finite $m$ where Alice and Bob stop after the first round which is a much more practical design. We discuss in detail those single-shot rates next.

\bigskip

\textbf{Single-shot purification.} Purification is complete after a single round if no photons are lost and Alice and Bob detect the same number of photons, $j_1=k_1$. If Bob detects less photons than Alice, $j_1<k_1$, further purification is required, however, it is most practical to disregard those outcomes and keep only the outcome $j_1=k_1$ which purifies in one shot. Alice's measurement can be preselected and prepared offline, thus, improving the rate of a particular pair of outcomes. The single-shot rate for a given $m$ and $k\equiv k_1 = j_1$ can be quite close to the PLOB rate. 

\rred{Recall that the optimal protocol implicit by the RCI is based on one-way classical communication~\cite{PhysRevLett.102.210501}, whereas, our iterative procedure requires two-way classical communication. Here, the single-shot protocol requires only one-way classical communication, like the implicit optimal protocol.} Alice and Bob agree on the quantum-error-detecting code ($k$ and $m$) in advance, so Alice does not need to send any classical information towards Bob. Bob only needs to send classical information towards Alice, telling her when the protocol succeeds. This greatly simplifies the required back-and-forth signalling in a quantum network.

\bigskip

\textbf{Highest achievable rate for single-shot purification against the bosonic pure-loss channel.} The rate of single-shot purification is as follows. The probability that Alice obtains outcome $k$ is $P_{k,m}^{\text{Alice}}=(1-\chi^2)^m \chi^{2k} d_{k,m}$, however, for the single-shot protocol we can incorporate this step into Alice's state preparation and assume this is done offline such that $P_{k,m}^{\text{Alice}}=1$. This means the rate will not depend on $\chi$. When Bob obtains the outcome which matches Alice's, $k$, his probability is $P_{k}^\text{Bob} = \eta^k$ and the output state is pure with entanglement $E_{k,m} = \log_2{d_{k,m}}$. Thus, the achievable rate of single-shot entanglement purification in ebits per use of the channel is
\begin{align}
    E_{k,m}^\text{single shot} (\eta)&= \frac{{P_{k}^\text{Bob} E_{k,m}}}{m}  = \frac{\eta^k }{m} {\log_2{ {k+m-1 \choose k}}}.\label{eq:rate}
\end{align}
The rate is divided by the number of rails, $m$, to compare directly with the PLOB bound~\cite{Pirandola_2017}. This is required because Alice and Bob exploit a quantum channel whose single use involves the simultaneous transmission of $m$ distinct systems in a generally entangled state.

This rate is for a perfect implementation without any additional losses, errors, or noise. The protocol heralds pure states in a single attempt, so if we have ideal quantum memories, after entanglement swapping (see Supplementary Note 4 for details on entanglement swapping) we can distribute entanglement between ends of a network without any loss of rate, i.e., at the rate given by~\cref{eq:rate} where $\eta$ is the transmissivity of the most destructive link.

Our optimised rate over $k$ and $m$ is shown in~\cref{fig:optimal_rate}a, without and with one quantum repeater (with a repeater, we assume ideal quantum memories). The PLOB bound can ultimately be broken at 46.3 km using single-shot purification between nodes and a single repeater for entanglement swapping. In~\cref{fig:optimal_rate}b, we show the ratio of our optimised single-shot rate with the PLOB bound, showing that at long distances the protocol falls short of the PLOB bound by just a factor of $\ln{3}/3 \sim 0.366$, and remarkably, our rate approaches $\frac{1}{2} C$ at short distances. At short distances, the rate is optimised for larger $k$ while at long distances $k=1$ is always optimal since the probability that photons arrive scales like $\eta^k$ at long distances. The optimal number of rails is about $m=3$ at most distances (including long distances) since increasing the number of rails decreases the rate like $1/m$. Larger $m$ is sometimes optimal at short distances.

The rate of our single-shot purification protocol is unable to saturate the ultimate limit because we postselect on Bob's outcomes, throwing away useful entanglement when $j\neq k$ and $j>0$. Keeping all measurement outcomes, the rate increases, however, it is less practical to do so.

\bigskip

\textbf{Quantum error detection.} In this section, we describe our single-shot purification protocol as quantum error detection. Consider encoding an arbitrary finite-dimensional single-rail state with dimension $d_{k,m}$:
\begin{align}
    \ket{\psi_{{k,m}}} &=  \frac{1}{\sqrt{N_\psi}} \sum_{\mu=0}^{d_{k,m}-1} c_{\mu} \ket{\mu},\label{eq:psi}
\end{align}
where $N_\psi=\sum_\mu^{d_{k,m}-1} |c_\mu|^2$.

The code is a subspace with $d_{k,m}$ dimensions (\mbox{$d_{k,m}$-dimensional}), a subspace of the infinite-dimensional Hilbert space chosen to detect loss errors nondeterminisitically. It is represented by the projector onto the subspace
\begin{align}
    \mathcal{P}_{k,m} = \sum_{\mu=0}^{{d_{k,m}}-1} \ketbra{\mu_{k,m}}{\mu_{k,m}},
\end{align}
where $ \ket{ \mu_{k,m}}= \ket{n_1^{(\mu)}, n_2^{(\mu)}, n_3^{(\mu)},\cdots,n_m^{(\mu)}}$ are the orthogonal basis states (code words) which make up the code subspace (code space), where $\mu$ is the logical label. Note $n_i^{(\mu)}$ is the number of photons in the $i$th rail, which depends on the logical label $\mu$, as well as $k$ and $m$ where $\sum_{i=1}^m n_i^{(\mu)}=k$. There are $d_{k,m}$ code words, i.e., for a given code the set of code words is $\{ \ket{ \mu_{k,m}} \} = \{ \ket{0_{k,m}}, \ket{1_{k,m}},\ket{2_{k,m}},\cdots \}$. Quantum information up to dimension $d_{k,m}$ can faithfully be transmitted to Bob, conditioned that he detects no errors. Photon loss (and photon gains) will result in states outside of code space, which we can distinguish as an error.

The set of logical states forming the \mbox{$d_{k,m}$-dimensional} basis of the code consists of all possible ways $k$ identical photons can be arranged among the $m$ distinct rails. For example, $\ket{\mu_{k=2,m=3}} \in \{\ket{0,0,2},\ket{0,2,0},\ket{2,0,0},\ket{0,1,1},\ket{1,0,1},\ket{1,1,0}\}$. The code space is a subspace of the full Hilbert space of the $m$ rails which introduces the redundancy required for error detection, that is, $k$ photons and $m$ rails can encode \mbox{$d_{k,m}$-dimensional} states. The dimension grows rapidly with $k$ and $m$. For example, with just $k=4$ photons and $m=5$ rails, we can efficiently encode 70-dimensional states, i.e., truncated at Fock number $\ket{69}$, with success probability $P_{j=k=4}^\text{Bob}=\eta^k=\eta^4$. This is a great advantage over noiseless linear amplification, for example, if we choose the gain of the amplifier to be $g=1/\sqrt{\eta}$ to overcome the loss then the success probability is  $P_\text{NLA}=1/g^{2(d_{k,m}-1)} = \eta^{(d_{k,m}-1)}=\eta^{69}$, totally impractical. Furthermore, noiseless linear amplification fails to purify completely and cannot completely overcome the loss.

The encoding step is
\begin{align}
    \mathcal{S}&= \sum_{\mu=0}^{d_{k,m}-1}  \ket{\mu_{k,m}} \bra{\mu},
\end{align}
which maps Fock states, $\ket{\mu}$, from a single mode to the code words, $\ket{\mu_{k,m}}=\ket{n_1^{(\mu)},n_2^{(\mu)},\cdots,n_m^{(\mu)}}$. The combined operation of encoding, loss, and decoding is a completely-positive trace-non-increasing map
\begin{align}
    \mathcal{E} &= \mathcal{S}^{-1} \circ  \mathcal{L}^{\otimes m} \circ \mathcal{S}=\mathcal{E} = \eta^{k/2} \sum_{\mu=0}^{d_{k,m}-1} \ketbra{\mu}{\mu},\label{eq:map}
\end{align}
where $ \mathcal{L}^{\otimes m}$ is the map for independent applications of the pure-loss channel on the $m$ rails, the decoding step, $\mathcal{S}^{-1}$, performs a QND measurement of the total photon number, and if $k$ photons arrive, then it successfully decodes back to single rail. The decoding step succeeds only if no photons are lost. The combined operation, $\mathcal{E}$, is a scaled identity map up to the $(d_{k,m}{-}1)$th Fock state, thus, the protocol succeeds with success probability $P_{k}^\text{Bob} = \tr{(\rho_{AB})} = \eta^k$ with unit fidelity.

The input state may be entangled with another mode. For example, we may consider encoding one arm of an arbitrary entangled state, $\ket{\psi_{k,m}} \propto  \sum_{\mu=0}^{d_{k,m}-1} c_\mu \ket{\mu}_A\ket{\mu}_B$. In this case, the operation, $\mathcal{E}$, acts on Bob's mode only and Alice leaves her mode alone. The final state shared between Alice and Bob is $\rho_{AB} = (\Id \otimes \mathcal{E}  ) \rho_{\text{in}}$, where $\rho_{\text{in}}$ is the initial state and $\Id$ is the identity on mode $A$. Note there is still useful entanglement if photons are lost. The entanglement-distribution rate of single-shot error detection for a maximally-entangled initial state $\ket{\phi_{k,m}}_{AB}=\frac{1}{\sqrt{{d_{k,m}-1}}} \sum_{\mu=0}^{d_{k,m}-1} \ket{\mu}_A\ket{\mu}_B$ is given by~\cref{eq:rate}, showing that error detection and purification indeed are equivalent.

One might consider using purification to distribute Gaussian entanglement for long-distance CV QKD between trusted end users of a network, performing the entanglement-based CV-QKD protocol based on homodyne detection~\cite{cerf2001quantum,gottesman2001secure} or heterodyne detection~\cite{weedbrook2004quantum}. Consider the initial data state to be a truncated TMSV state with dimension $d_{k,m}$ given by
\begin{align}
    \ket{\chi_{k,m}} &=  \sqrt{\frac{\chi^2-1}{\chi^{2{d_{k,m}}}-1}} \sum_{\mu=0}^{d_{k,m}-1} \chi^\mu \ket{\mu}_A \ket{\mu}_B.\label{eq:chi}
\end{align}
The state is Gaussian except for the hard truncation in Fock space. The protocol works as follows. First, truncated TMSV states, $\ket{\chi_{k,m}}$, are distributed using our single-shot purification scheme between all repeater nodes and held in quantum memories. Once successful, CV entanglement swapping is used to entangle the end users who use the entanglement to perform CV QKD. While the purification scheme can be complex (depending on the chosen protocol size), the CV entanglement swapping is simple. It works by performing dual-homodyne measurements on some of the modes, followed by conditional displacements, swapping the entanglement~\cite{PhysRevA.83.012319}, see Supplementary Note 4 for details on how to compute the secret key rate.

\bigskip

\textbf{A simple example: the qubit code.} The simplest nontrivial code uses a single photon, $k=1$, in two rails, $m=2$, (i.e., unary dual-rail) and protects qubit systems, $d_{k=1,m=2}=2$, from loss. It is equivalent to the original purification protocol from Ref.~\cite{Bennett_1996}, but in this context we use it to purify entanglement completely from pure loss to first order in Fock space and we can protect arbitrary single-rail qubit states from loss. The code words can be defined $\ket{{0}_{k=1,m=2}} \equiv \ket{0,1}$ and $\ket{{1}_{k=1,m=2}} \equiv \ket{1,0}$. The projector onto the code space is $\mathcal{P} =\ketbra{0,1}{0,1} + \ketbra{1,0}{1,0}$. The encoding step is
\begin{align}
    \mathcal{S} &= \ketbra{{0}_{k=1,m=2}}{0} + \ketbra{{1}_{k=1,m=2}}{1}\\
    &= \ketbra{0,1}{0} + \ketbra{1,0}{1},
\end{align}
which maps the vacuum component of the data mode onto a single photon of the second rail and the single-photon component of the data mode onto a single photon of the first rail. If either of these photons is lost, the protocol fails. The success probability is $P_{j=k=1}^\text{Bob}=\eta$. The maximum single-shot rate in ebits per use is $D_{k{=}1,m{=}2}^\text{single shot}(\eta)=\eta/2$, as shown by the dashed line in~\cref{fig:optimal_rate}a (with and without a repeater).

\bigskip

\textbf{Physical implementation.} Entanglement purification (iterative and single shot) requires joint QND measurements on multiple rails. These measurements can be performed via controlled-SUM quantum gates and photon number-resolving measurements (see Supplementary Note 5, for example). Another technique is to use high finesse cavities and cross–Kerr nonlinearities~\cite{PhysRevLett.84.4002}. 

\bblack{More simply, our scheme can be implemented using beamsplitters and photon-number-resolving detectors, however, it also requires an entangled resource state, $\ket{\Omega_{k,m}}$. This is a common technique in linear optics~\cite{Yan:21,Yan2021}. See Ref.~\cite{Kok2007} for a review of quantum information processing using linear optics. The resource state can also be generated using linear optics and photon-number measurements. That is, we have a simple method of implementing our protocol, though at some cost to the success probability.}

\begin{figure}
\includegraphics[width=1\linewidth]{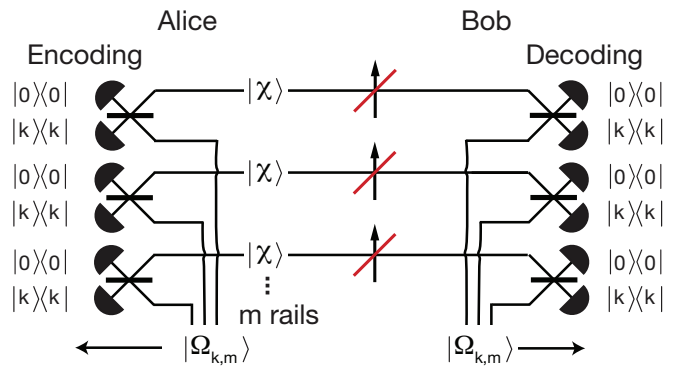}
\caption{\bblack{\textbf{Implementation using linear optics and number measurements.}} The aim of the protocol is for Alice to share quantum information contained in a \mbox{$d_{k,m}$-dimensional} pure state, $\ket{\psi_{k,m}}$, which can be entangled with another system which Alice keeps, with Bob separated by a pure-loss channel. Shown in this figure, the aim is to distribute a pure maximally-entangled state between Alice and Bob. Alice shares $m$ entangled states, $\ket{\chi}$, with Bob. Encoding/decoding simply requires beamsplitters, photon detectors, and $(m+1)$-mode entangled resource states, $\ket{\Omega_{k,m}}$. We can assume Alice's encoding (the state heralded after Alice's measurement) is prepared offline so that the success probability of the protocol is the probability of Bob's side only. There is, however, a success-probability penalty for using linear optics, noting crucially that this penalty does not depend on the loss. The protocol succeeds when Bob detects the same number of photons Alice sent, $k$. If Bob detects less photons, multiple rounds of error detection are required or Alice and Bob can simply disregard those states.}
\label{fig:implementation}
\end{figure}

We focus on the single-shot linear-optics protocol, shown in~\cref{fig:implementation}. Alice shares $m$ entangled states, $\ket{\chi}$, with Bob\bblack{. Note each rail could have a different $\chi$, distilling different amounts of entanglement between Alice and Bob at the end of the protocol, however, maximal entanglement is heralded when all rails have the same value of $\chi$.} \bblack{Alice and Bob each prepare locally multimode resource states, $\ket{\Omega_{k,m}}$, consisting of $(m+1)$ modes. We assume they do this offline so it does not affect the quantum communication rate:}
\begin{align}
    \ket{\Omega_{k,m}} \propto \sum_{\mu=0}^{d_{k,m}-1} f_\mu \ket{\mu} \ket{{\mu_{\widetilde{k,m}}}},\label{eq:omega}
\end{align}
where $\ket{\mu}$ are Fock states, $\ket{{\mu_{\widetilde{k,m}}}}$ are ``anticorrelated'' code words. Writing the usual code words in the Fock basis as $\ket{{\mu_{k,m}}}= \ket{n_1^{(\mu)}, n_2^{(\mu)}, n_3^{(\mu)},\cdots,n_m^{(\mu)}}$, where $n_i^{(\mu)}$ is the number of photons in the $i$th rail, and where $\sum_{i=1}^m n_i^{(\mu)}=k$, recalling that the code space was defined as the set of all states with this property, then $\ket{{\mu_{\widetilde{k,m}}}} =  \ket{k{-}n_1^{(\mu)}, k{-}n_2^{(\mu)}, k{-}n_3^{(\mu)},\cdots,k{-}n_m^{(\mu)}}$. The coefficients, $f_n$, are
\begin{align}
     f_n &= \left [ {k \choose n_1^{(\mu)}} {k \choose n_2^{(\mu)}} \cdots {k \choose n_m^{(\mu)}} \right]^{-1/2}.\label{eq:fn}
\end{align} 

\bblack{The last $m$ modes of  \ket{\Omega_{k,m}} are fed into the beamsplitters with the distributed entanglement and are measured by photon-number detectors. The first mode is kept locally by the user and remains at the end of the protocol, as shown in ~\cref{fig:implementation}.}

Detecting $k$ photons means there were no loss events since $n_i^{(\mu)} + k - n_i^{(\mu)} = k$. That is, all photons are accounted for in the circuit and the output state is strictly pure. There may be useful entanglement for measurement outcomes other than $\ket{0,k}\bra{0,k}$ at each pair of detectors, and further purification (iteration) can increase the rate, but we do not consider it due to practicality.

Adjusting the amount of entanglement prepared for each rail, adjusting the loss on each rail, or selecting different coefficients in the resource state, $f_n$, results in a different output state, which may be useful for certain tasks. For example, if the entangled states prepared for each rail are identical and have the same amount of squeezing and $f_n$ is chosen as in~\cref{eq:fn}, then a maximally-entangled state will be heralded between Alice and Bob at the output. For another example, consider the dual-rail case ($m=2$), if the second rail is maximally entangled and $f_n$ is chosen as in~\cref{eq:fn}, then the output state is the initial state of the top rail. This tuning of the circuit parameters is useful, for instance, for CV QKD where the target state is a truncated TMSV state.

The linear-optics scheme detects all errors and outputs a pure state. There is, however, an additional success probability penalty using linear optics because of Bob's decoding measurement. We assume Alice prepares offline, then the success probability is
\begin{equation}
    P_{k,m}^\text{Bob linear optics} = \frac{ \eta^k}{2^{m(k-1)}\sum_{n=0}^{d-1} f_n^2},\label{eq:P}
\end{equation}
which depends on $m$. Compare this with the ideal purification protocol where $P_{k}^\text{Bob}=\eta^k$ which does not depend on $m$. The penalty paid for using linear optics is mainly due to the exponential $2^{m(k-1)}$ factor, which is painful for anything other than $k=1$, where $P_{k=1}^\text{Bob linear optics}=\eta/m$. For more details we refer you to Supplementary Note 6.

The linear-optics circuit leads naturally to a controlled-SUM gate using linear optics \bblack{and number measurements}, albeit with distorted coefficients (this distortion ultimately has no effect in our protocol since we immediately measure the state). For the interested reader, we refer you to Supplementary Note 5 for more details.

Since Alice's encoded state is prepared offline, it is useful to consider more generally that she encodes an arbitrary state into the code: $(\Id \otimes \mathcal{S}) \ket{\psi_{k,m}}$.

\bigskip

\textbf{Preparation of resource states.} For our \bblack{linear-optics and number measurement} circuit, the resource state, $\ket{\Omega_{k,m}}$, and Alice's encoded state she prepares offline, are multi-mode entangled states. Once these states are prepared, our scheme requires just beamsplitters and photon detectors. One practical way to prepare these states is to use a Gaussian Boson Sampler (GBS)~\cite{PhysRevLett.119.170501} and postselecting on a specific photon-number-resolving measurement click pattern on some of the modes of the output~\cite{PhysRevA.100.052301,PhysRevA.100.022341}. This allows our scheme to be implemented entirely using linear-optics \bblack{and number measurements}. Using the GBS method for the simplest scheme with $k=1$ and $m=2$, we have found the resource state, $\ket{\Omega_{k{=}1,m{=}2}} = (\ket{0,1,0} + \ket{1,0,1}) /\sqrt{2}$, can be prepared with high fidelity, $F>0.999$, with success probability $\approx 10^{-6}$. This was found by optimising the parameters of a GBS network via a machine learning algorithm called ``basin hopping''~\cite{PhysRevA.100.012326}. In~\cite{Josh_github}, we have provided our code which implements this algorithm, as well as the parameter set we found that generates this resource state. Alternatively, adaptive phase measurements~\cite{Lund_2005} can be used to prepare the needed resource states directly from dual-rail Bell pairs or GHZ-like states which may have a higher probability of success.

\bigskip

\textbf{Experimental imperfections.} Our linear-optics circuit is robust to loss. The quality of the state Alice sends can be managed since her encoding is done offline. In any case, we are interested in the noise introduced by the protocol, not the noise in the initial state we are trying to transmit. The measurement and detection scheme at Bob's side is such that all photons are accounted for, so if Alice's encoding is perfect, the protocol can correctly identify if any photons are lost in the channel or at the detectors. More details are presented in  Supplementary Note 6, \bblack{where we also perform a numerical simulation incorporating inefficient detectors with dark counts, and a thermal-noise channel. Our protocol is robust to practical values of these imperfections.}

\bigskip

\textbf{The directionality issue.} Often in quantum communications it is best if quantum states propagate in one preferred direction (from Alice towards Bob) when reverse reconciliation is used. This is a persistent problem in CV quantum communications. For instance, CV measurement-device-independent protocols~\cite{Ma_2014,Zhang_2014,Ottaviani_2015,Pirandola2015} work only in an extremely asymmetric configuration, with the node (ineffective as a repeater) positioned close to one of the trusted parties. This directionality problem is also present in the repeater protocols considered in Ref~\cite{Dias_2020,Ghalaii_2020}, where they work best in one direction, which for reverse reconciliation is again from Alice towards Bob. Our purification schemes allow states to propagate in any direction in a quantum network since the output states are pure. Using purification for CV-QKD works the same for both direct and reverse reconciliations. We have no directionality problem. This is an important requirement for large CV networks.  Note that the memoryless CV repeater protocol introduced in Ref.~\cite{winnel2021overcoming} also fixes the directionality problem.

\section*{Discussion}\label{sec:discussion}

\bblack{We have presented a physical protocol which achieves the two-way assisted quantum capacity of the pure-loss channel~\cite{Pirandola_2017}.} Error \textit{correction} requires short distances between neighbouring repeater nodes, while in contrast, we showed simple error \textit{detection} can saturate the PLOB bound. An open question is what protocol can saturate the fundamental limits for added thermal noise and close the gap between the theoretical upper and lower bounds~\cite{Pirandola_2017}. Furthermore, can error detection saturate the ultimate limits for other noise models, \bblack{for instance, dephasing?}

\bblack{Our protocol is an optimal one and can be performed using CSUM quantum gates and measurements. However, it is unknown whether other gates and measurements may also achieve the capacity, and if they can do so more efficiently, i.e., approaching the PLOB bound more quickly with the number of iteration rounds, $n$.}

Our ultimate protocol is experimentally challenging since it requires iterative purification steps. However, we show that by simplifying our protocol to just a single round of purification, we can still achieve excellent rates. This is superior to other nondeterministic techniques such as noiseless linear amplification where it is impossible to remove all effects of loss.

Our purification protocol can be implemented using \bblack{linear-optics and number measurements}. This does introduce some probability penalty (which does not depend on the loss), however, having an all-optical design is experimentally convenient.

One limitation of our results is the requirement for high-performance quantum memories. However, purification outputs pure states which is extremely beneficial. Firstly, pure states can handle a higher amount of decoherence coming from nonideal quantum memories. Secondly, distributing pure states will prevent how much thermal noise builds up across a network during entanglement swapping. Finally, purity may be beneficial for both point-to-point and repeater-assisted CV QKD, improving the signal-to-noise ratio and decreasing Eve's knowledge of the key, speeding up the classical post-processing part of the protocol. 

Thus, purification is inherently a useful technique for overcoming the unavoidable losses in quantum communication networks. Remarkably, purification can be employed totally using linear optics \bblack{and number measurements} and is compatible with existing DV and CV infrastructure. Finally, we remark that since the lossy entanglement is completely purified, it can be used for exotic tasks, such as device-independent quantum key distribution and demonstrating Bell nonlocalities, over much longer distances than previously imagined.

\section*{Methods\label{sec:methods}}

\bigskip

\textbf{Noise model.} Bosonic pure loss is the dominant source of noise for many quantum communication tasks. The pure-loss channel is equivalent to introducing a vacuum mode and mixing the data state on a beamsplitter with transmissivity $\eta \in [0,1]$. The Kraus-operator representation of the single-mode pure-loss channel is~\cite{PhysRevA.56.1114}
\begin{align}
    \mathcal{L} (\rho) &= \sum_{l=0}^{\infty} A_l \rho A_l^\dagger,\label{eq:pure_loss_Kraus_represenation}
\end{align}
with Kraus operators $A_l  = \sqrt{\frac{(1-\eta)^l}{l!}}\eta^{\hat{\frac{n}{2}}}\hat{a}^l$ associated with losing $l$ photons to the environment, where $\hat{a}$ and $\hat{a}^\dagger$ are the single-mode annihilation and creation operators, respectively, and $\hat{n}=\hat{a}^\dagger \hat{a}$ is the photon-number operator.

\bigskip

\textbf{Reverse coherent information.} Take a maximally entangled state of two systems $A$ and $B$. Propagating the $B$ system through the quantum channel defines the Choi state of the channel. Then the reverse coherent information represents a lower
bound for the distillable entanglement and for the optimal secret key rate. The reverse coherent information of a state $\rho_{AB}$ is defined as~\cite{PhysRevLett.102.210501}
\begin{align}
I(\rho_{AB}) &= S(\rho_A) - S(\rho_{AB}),
\end{align}
where $S(\rho_A)$ and $S(\rho_{AB})$ are von Neumann entropies of $\rho_A = \tr_B(\rho_{AB})$ and $\rho_{AB}$ respectively. The von Neumann entropy of $\rho$ is $-\tr(\rho \log_2 \rho)$.



\bigskip

\begin{acknowledgements}

We thank Ozlem Erkilic, Sebastian Kish, Ping Koy Lam, Syed Assad, \rred{Deepesh Singh}, and Josephine Dias for valuable discussions during this investigation. \rred{We thank Deepesh Singh for an alternative proof given in Supplementary Note 7.} This research was supported by the Australian Research Council (ARC) under the Centre of Excellence for Quantum Computation and Communication Technology (CE170100012).

\end{acknowledgements}

\section*{Author contributions}
All authors contributed to this work extensively and to the writing of the manuscript.

\section*{Competing Interests statement}
All authors reported no potential competing interests.

\bblack{
\section*{Data availability statement}
The datasets generated during and analysed during the current study are available from the corresponding author on reasonable request.
}

%

\newcounter{S} \setcounter{section}{0} \setcounter{subsection}{0}
\setcounter{figure}{0}\setcounter{equation}{0}
\renewcommand{\thefigure}{\arabic{figure}} \renewcommand{\figurename}{Supplementary Figure}

\renewcommand{\theequation}{S\arabic{equation}}
\renewcommand{\bibnumfmt}[1]{[S#1]} {} \renewcommand{\citenumfont}[1]{S#1}

\renewcommand{\thesection}{Supplementary Note \arabic{section}} \renewcommand\refname{Supplementary References}

\clearpage
\onecolumngrid
\bigskip

\bigskip

\begin{center}
{\Large Supplementary Notes}
\end{center}

\crefname{figure}{Supplementary Fig.}{Supplementary Figs.}
\Crefname{figure}{Supplementary Figure}{Supplementary Figures}

\section{Noiseless linear amplification and entanglement swapping cannot completely purify entanglement}\label{app:NLA}

Noiseless linear amplification is a distinguished technique for the distillation of continuous-variable (CV) entanglement but it is unable to purify completely from loss. It is unable to return a loss-attenuated entangled state back to its original loss-free entangled state.

To see this, consider pure loss acting on one mode of a CV two-mode squeezed vacuum (TMSV) state with the squeezing parameter $\chi\in [0,1]$. Pure loss is equivalent to interacting the data mode with vacuum on a beamsplitter with transmissivity $\eta\in[0,1]$. Consider ideal noiseless linear amplification~\cite{SUP_Pandey_2013} with gain $g\geq1$ applied to the lossy arm to correct the loss. The final state after ideal noiseless linear amplification of the lossy mode results in another {\it{lossy mixed}} TMSV state but with different parameters:
\begin{align}
    \ket{\chi} \lossto \mathcal{L}_\eta (\ketbra{\chi}{\chi}) \NLAto \mathcal{L}_{\eta'} (\ketbra{\chi'}{\chi'}),
\end{align}
where $\mathcal{L}_\eta$ is the completely-positive trace-preserving (CPTP) map for the pure-loss channel with transmissivity $\eta$. The new parameters, $\chi'$ and $\eta'$, are related to the old parameters, $\chi$ and $\eta$, in the following way:~\cite{SUP_PhysRevA.86.012327, SUP_PhysRevA.89.023846, SUP_PhysRevA.91.062305}
\begin{align}
    \chi' &= \sqrt{1-\eta+\eta g^2} \chi,\\
    \eta' &= \frac{g^2}{1-\eta+\eta g^2} \eta.
\end{align}
The effective loss is zero when $\eta'=1$ which for finite gain is only possible when $\eta=1$. This means that it is impossible to completely correct loss using noiseless linear amplification. Also, as $g\to\infty$, the success probability tends to zero.

Entanglement swapping is another technique that can distil entanglement, however, it can be thought of as noiseless linear amplification with some loss distributed across the measurement~\cite{SUP_winnel2021overcoming,SUP_guanzon2021ideal}, thus, entanglement swapping is also unable to completely purify CV entanglement. Other distillation techniques such as photon subtraction are also strictly limited in their purification abilities~\cite{SUP_subtraction2010}.

Noiseless linear amplification is nondeterministic. Ideal noiseless linear amplification is unphysical since the success probability is zero. Physical noiseless linear amplification is possible with a non-zero success probability by considering an energy cutoff. For instance, consider noiseless linear amplification which amplifies and truncates the input state at Fock number $d-1$ (where $d$ is the dimension of the output state), then the success probability scales like $\eta^{d-1}$. Our purification scheme uses the entanglement between multiple modes so larger states can be protected against loss with a much higher success probability than noiseless linear amplification. Our success probability depends on the number of photons in the code, not the dimension of the state that is encoded. Our single-shot purification scheme has the improved success probability $\eta^{k}$ where in general $k\leq d-1$. Furthermore, purification removes completely the effect of pure loss since errors are detected, while noiseless linear amplification is unable to do this.

To summarise, noiseless linear amplification and entanglement swapping (i.e., techniques for the first and second generations of quantum repeaters) fail to saturate the ultimate rate limits under pure loss for two reasons: 1. the output entangled states are strictly non-pure for any amount of loss and for finite gain, and 2. the success probability depends on the energy cutoff and tends to zero for large input states. Whereas, our purification scheme purifies completely and has a success probability that scales optimally with loss in the sense that the highest average rate of entanglement purification saturates the capacity (the PLOB bound~\cite{SUP_Pirandola_2017}).

\section{Derivation of purified states generated between Alice and Bob}\label{app:derivation_states}

In this Supplementary Note, we derive the purified states heralded by Alice's and Bob's QND measurements.

\subsection*{Heralded states after round one of purification} 

In this section, we derive the heralded states (equation (3) of the main text) during round one.

Alice prepares $m$ copies of infinite-dimensional TMSV states, $\ket{\chi}$, where $\chi \in [0,1]$ is the squeezing parameter. Then, she performs a mode-blind QND measurement on her side, and obtains an outcome of $k_1$ total photons. This measurement projects the $m$ TMSV states onto a maximally-entangled state, $\ket{\phi_{k_1,m}}_{AB}$, with dimension $d_{k_1,m} = {k_1+m-1 \choose k_1}$, entanglement $E_{k,m} = \log_2{d_{k,m}}$, and success probability $P_{k_1,m}^{\text{Alice}} = (1-\chi^2)^m \chi^{2 k_1} d_{k_1,m}$. After Alice obtains outcome $k_1$, Alice has the following maximally-entangled state:
\begin{align}
\ket{\phi_{k_1,m}}_{AB}=(1-\chi^2)^\frac{m}{2} \chi^{k_1}  \stackrel{n_{1}+n_{2}+\cdots +n_{m}=k_1}{\mathrel{\mathop{\sum }\limits_{n_{1},n_{2},\cdots ,n_{m}}}}   \ket{ n_{1},n_{2},\cdots ,n_{m}} _{A} \ket{ n_{1},n_{2},\cdots ,n_{m}}_{B },
\end{align}
where $A$ refers to the quantum system Alice keeps and $B$ refers to the quantum system which will be sent to Bob.

Alice propagates Bob's modes ($B_1,B_2,...,B_m$) across independent pure-loss channels. The pure loss is equivalent to introducing $m$ vacuum modes and mixing those with the $m$ data rails on beamsplitters of transmissivity $\eta$. We label the beamsplitter transformations $T_i$ which interact modes $B_i$ and $e_i$. The state after this interaction is
\begin{align}\label{eq:after_channel}
T_1 T_2 T_3 \cdots T_m \ket{\phi_{k_1,m}}_{AB} \ket{0}_{e}  &= (1-\chi^2)^\frac{m}{2} \chi^{ k_1} \stackrel{n_{1}+n_{2}+\cdots +n_{m}=k_1}{\mathrel{\mathop{\sum }\limits_{n_{1},n_{2},\cdots ,n_{m}}}}   \stackrel{n_1}{\mathrel{\mathop{\sum }\limits_{l_1=0}}} \; \stackrel{n_2}{\mathrel{\mathop{\sum }\limits_{l_2=0}}} \; \stackrel{n_3}{\mathrel{\mathop{\sum }\limits_{l_3=0}}} \cdots \stackrel{n_m}{\mathrel{\mathop{\sum }\limits_{l_m=0}}} \; \sqrt{{n_1 \choose l_1}  {{n_2 \choose l_2}} {{n_3 \choose l_3}} \cdots {{n_m \choose l_m}}}  \; \\ &\;\;\;\;\;\; (1-\eta)^\frac{l_1+l_2+\cdots+l_m}{2} \eta^\frac{n_1+n_2+\cdots+n_m - (l_1+l_2+\cdots+l_m)}{2} \nonumber \\
&\;\;\;\;\;\;\;\;\;  \nonumber \ket{ n_{1},n_{2},\cdots ,n_{m}} _{A }  \ket{ n_{1}-l_{1},n_{2}-l_{2},\cdots ,n_{m}-l_{m}}_{B } \ket{ l_{1},l_{2},\cdots ,l_{m}}_{e },
\end{align}
where $l_i$ is the number of photons lost from mode $B_i$ to $e_i$.

Finally, Bob performs a QND measurement and obtains an outcome of $j_1$ total photons. The terms that survive in~\cref{eq:after_channel} are those components where Bob's modes have total photons which add up to $j_1$, that is, $n_1-l_1+n_2-l_2+\cdots+n_m-l_m=j_1$. The total number of lost photons is the difference between Alice and Bob's outcomes, $l_1+l_2+\cdots+l_m = k_1-j_1$. Thus, the unnormalised final state shared between Alice and Bob is
\begin{align}
\ket{\phi_{k_1,j_1,m}}_{ABe }  &= (1-\chi^2)^\frac{m}{2} \chi^{ k_1} (1-\eta)^\frac{k_1-j_1}{2} \eta^\frac{j_1}{2} \stackrel{n_{1}+n_{2}+\cdots +n_{m}=k_1}{\mathrel{\mathop{\sum }\limits_{n_{1},n_{2},\cdots ,n_{m}}}} \;  \nonumber \stackrel{l_1+l_2+\cdots+l_m=k_1-j_1,\; l_i\leq n_i \forall i}{\mathrel{\mathop{\sum }\limits_{l_{1},l_{2},\cdots ,l_{m}}}} \;  \nonumber \; \sqrt{{n_1 \choose l_1}  {{n_2 \choose l_2}} {{n_3 \choose l_3}} \cdots {{n_m \choose l_m}}} \; \\
&\;\;\;\;\;\;\;\;\;   \ket{ n_{1},n_{2},\cdots ,n_{m}} _{A}  \ket{ n_{1}-l_{1},n_{2}-l_{2},\cdots ,n_{m}-l_{m}}_{B } \ket{ l_{1},l_{2},\cdots ,l_{m}}_{e  },
\end{align}
which is equation (3) of the main text as required. 

We can calculate the success probability (i.e., the probability to successfully get this output state) as follows
\begin{align}
    P_{k_1,j_1,m} &= \langle \phi_{k_1,j_1,m} | \phi_{k_1,j_1,m} \rangle \\ 
    &= (1-\chi^2)^m \chi^{2 k_1} (1-\eta)^{k_1-j_1} \eta^{j_1} \stackrel{n_{1}+n_{2}+\cdots +n_{m}=k_1}{\mathrel{\mathop{\sum }\limits_{n_{1},n_{2},\cdots ,n_{m}}}} \;  \stackrel{l_1+l_2+\cdots+l_m=k_1-j_1,\; l_i\leq n_i \forall i}{\mathrel{\mathop{\sum }\limits_{l_{1},l_{2},\cdots ,l_{m}}}} \; \; {{n_1 \choose l_1}  {{n_2 \choose l_2}} {{n_3 \choose l_3}} \cdots {{n_m \choose l_m}}} \; \\
    &= (1-\chi^2)^m \chi^{2 k_1} (1-\eta)^{k_1-j_1} \eta^{j_1} \stackrel{n_{1}+n_{2}+\cdots +n_{m}=k_1}{\mathrel{\mathop{\sum }\limits_{n_{1},n_{2},\cdots ,n_{m}}}} {k_1 \choose j_1} \\ 
    &= (1-\chi^2)^m \chi^{2k_1} d_{k,m} (1-\eta)^{k_1-j_1} \eta^{j_1} {k_1 \choose j_1}\\
    &= P_{k_1,m}^{\text{Alice}} P_{k_1,j_1}^\text{Bob},
\end{align}
We used the generalized Vandermonde identity to simplify the inner summation to ${k_1 \choose j_1}$; this does not depend on the outer summation, which has $d_{k_1,m}={k_1+m-1 \choose k_1}$ terms. We can assign $P_{k_1,m}^{\text{Alice}}=(1-\chi^2)^m \chi^{2k_1} d_{k,m}$ and $P_{k_1,j_1}^\text{Bob}=(1-\eta)^{k_1-j_1} \eta^{j_1} {k_1 \choose j_1}$, where we can interpret $P_{k_1,j_1,m}=P_{k_1,m}^{\text{Alice}} P_{k_1,j_1}^\text{Bob}$ as a joint probability distribution. Remarkably, while this success probability $P_{k_1,j_1,m}$ depends on the transmissivity $\eta$, the renormalised output state 
\begin{align}
    \ket{\phi_{k_1,j_1,m}'} = \ket{\phi_{k_1,j_1,m}}/\sqrt{P_{k_1,j_1,m}},
\end{align}
shared between Alice and Bob does not depends on $\eta$.

\subsection*{Round two of purification}

Recall that in the first round, Alice obtains outcome $k_1$ and Bob obtains outcome $j_1$. They share the following state:
\begin{align}
\ket{\phi_{k_1,j_1,m}}_{ABe }  &= (1-\chi^2)^\frac{m}{2} \chi^{ k} (1-\eta)^\frac{k_1-j_1}{2} \eta^\frac{j_1}{2} \stackrel{n_{1}+n_{2}+\cdots +n_{m}=k_1}{\mathrel{\mathop{\sum }\limits_{n_{1},n_{2},\cdots ,n_{m}}}} \;  \nonumber \stackrel{l_1+l_2+\cdots+l_m=k_1-j_1,\; l_i\leq n_i \forall i}{\mathrel{\mathop{\sum }\limits_{l_{1},l_{2},\cdots ,l_{m}}}} \;  \nonumber \; \sqrt{{n_1 \choose l_1}  {{n_2 \choose l_2}} {{n_3 \choose l_3}} \cdots {{n_m \choose l_m}}} \; \\
&\;\;\;\;\;\;\;\;\;   \ket{ n_{1},n_{2},\cdots ,n_{m}} _{A }  \ket{ n_{1}-l_{1},n_{2}-l_{2},\cdots ,n_{m}-l_{m}}_{B } \ket{ l_{1},l_{2},\cdots ,l_{m}}_{e}.
\end{align}
Let us firstly consider the case when $j_1=k_1$, then $l_1+l_2+\cdots+l_m=0$ thus the output state is:
\begin{align}
\ket{\phi_{k_1,k_1,m}}_{ABe}  &= (1-\chi^2)^\frac{m}{2} \chi^{ k_1} \eta^\frac{k_1}{2} \stackrel{n_{1}+n_{2}+\cdots +n_{m}=k_1}{\mathrel{\mathop{\sum }\limits_{n_{1},n_{2},\cdots ,n_{m}}}}   \ket{ n_{1},n_{2},\cdots ,n_{m}} _{A }  \ket{ n_{1},n_{2},\cdots ,n_{m}}_{B } \ket{0,0,\cdots,0}_{e}.
\end{align}
We can see that the environment modes are all vacuum states and are separable from the state. When $j_1=k_1$ the state is purified in a single shot, and we do not need a second round.

Now, consider the case when $j_1<k_1$, we will need further rounds of purification. In the second round, one option for further purification is to measure the total photon number on $m-1$ rails instead of all $m$ rails. This means they learn if some errors happened in the $m$th rail, or in the first $m-1$ rails. If Alice obtains outcome $k_2$ and Bob obtains outcome $j_2$, they share the state:
\begin{align}
\ket{\phi_{k_1,j_1,k_{2}, j_{2},m}}_{ABe }  &= (1-\chi^2)^\frac{m}{2} \chi^{ k_1} (1-\eta)^\frac{k_1-j_1}{2} \eta^\frac{j_1}{2} \nonumber \\
&\;\;\; \stackrel{n_{1}+n_{2}+\cdots +n_{m}=k_1,\; n_{1}+n_{2}+\cdots +n_{m-1}=k_2}{\mathrel{\mathop{\sum }\limits_{n_{1},n_{2},\cdots ,n_{m}}}} \;\;\; \nonumber \stackrel{l_1+l_2+\cdots+l_m=k_1-j_1,\; l_1+l_2+\cdots+l_{m-1}=k_2-j_2,\; l_i\leq n_i \forall i}{\mathrel{\mathop{\sum }\limits_{l_{1},l_{2},\cdots ,l_{m}}}} \;  \nonumber \;\\
&\;\;\sqrt{{n_1 \choose l_1}  {{n_2 \choose l_2}} {{n_3 \choose l_3}} \cdots {{n_m \choose l_m}}} \;   \ket{ n_{1},n_{2},\cdots ,n_{m}} _{A }  \ket{ n_{1}-l_{1},n_{2}-l_{2},\cdots ,n_{m}-l_{m}}_{B } \ket{ l_{1},l_{2},\cdots ,l_{m}}_{ e}.
\end{align}

If $j_1<k_1$ and $j_2=k_2$, then no errors occurred in the first $m-1$ rails and the  output state is $\propto \stackrel{n_{1}+n_{2}+\cdots +n_{m-1}=k_2,\; n_{m}=k_1-k_2}{\mathrel{\mathop{\sum }\limits_{n_{1},n_{2},\cdots ,n_{m}}}}  \ket{ n_{1},n_{2},\cdots ,n_{m-1},k-k_2} _{A } \ket{ n_{1},n_{2},\cdots ,n_{m-1},j-j_2}_{B } \ket{ 0,0,\cdots ,0, k_1-j_1}_{e }$, so the $m$th rail is in a definite quantum state with no entanglement so this rail should be discarded. The rest of the $m-1$ modes are not entangled with the environment, so Alice and Bob's state of these rails is pure.

Finally, if $j_1<k_1,j_2<k_2$, we need still further purification rounds. After $m-1$ iterative rounds, we show next that we achieve the capacity as $\chi\to1,m\to\infty$.

\section{Optimality of our protocol and rates}\label{app:optimality}

In this Supplementary Note, we show that our protocol is optimal for entanglement distillation and achieves the two-way assisted capacity of the pure-loss channel (the PLOB bound) as $m\to\infty$ and $\chi\to1$.

\subsection*{Proof that the first round is optimal}

In this section, we show that the first round is optimal in the limit that $m\to\infty$ and provide a lower bound on the distillable entanglement by directly calculating the average reverse coherent information (RCI).

The RCI gives a lower bound on the distillable entanglement and is given by (see Methods and Ref.~\cite{SUP_PhysRevLett.102.210501})
\begin{align}
R(\rho_{AB}) &= S(\rho_A) - S(\rho_{AB}),\label{eq:RCI}
\end{align}
where $S(\rho_A)$ and $S(\rho_{AB})$ are the von Neumann entropies of $\rho_A = \tr_B(\rho_{AB})$ and $\rho_{AB}$ respectively, where $S(\rho)=-\tr(\rho \log_2 \rho)$.

For our iterative purification protocol, the average RCI per channel use is
\begin{align}
     S_1+F_1 &= \frac{1}{m} \sum_{k_1=0}^\infty \sum_{j_1=0}^{k_1} P_{k_1,j_1,m} R_{k_1,j_1,m},
\end{align}
where $P_{k_1,j_1,m}$ is the probability of Alice and Bob obtaining outcomes $k_1$ and $j_1$, heralding a state with RCI, $R_{k_1,j_1,m}$. $S_1$ is the average entanglement of the pure states where the RCI is equal to the von Neumann entropy (when $k_1=j_1$):
\begin{align}
     S_1 &= \frac{1}{m} \sum_{k_1=0}^\infty \sum_{j_1=0}^{k_1} P_{k_1,j_1,m} R_{k_1,j_1,m} \rred{\delta_{k_1,j_1} = \frac{1}{m} \sum_{k_1=0}^\infty P_{k_1,j_1=k_1,m} R_{k_1,j_1=k_1,m} }, \label{eq:S_}
\end{align}
and $F_1$ is the average RCI of the failed terms (when $k_1 \neq j_1$):
\begin{align}
     F_1 &= \frac{1}{m} \sum_{k_1=0}^\infty \sum_{j_1=0}^{k_1} P_{k_1,j_1,m} R_{k_1,j_1,m} \rred{(1-\delta_{k_1,j_1}) = \frac{1}{m} \sum_{k_1=0}^\infty \sum_{j_1=0}^{k_1-1} P_{k_1,j_1,m} R_{k_1,j_1,m} }. \label{eq:F_}
\end{align}
The sum over $j_1$ terminates at $k_1$ since Bob cannot obtain outcomes $j_1>k_1$. \rred{Notice the Kronecker delta $\delta_{k_1,j_1}$ in these equations highlights that the only difference between~\cref{eq:S_} and~\cref{eq:F_} is the constraint: $j_1=k_1$ for S18 and $j_1 \neq k_1$ for S19. That way, it is easy to see that the total weighted average rate of success and failure is $S_1+F_1=\frac{1}{m} \sum_{k_1=0}^\infty \sum_{j_1=0}^{k_1} P_{k_1,j_1,m} R_{k_1,j_1,m}$, without constraint.}

We calculate the RCI, $R_{k_1,j_1,m}$, of the normalised state $\rho_{AB} = \Tr_E(\ket{\phi_{k_1,j_1,m}'}{\bra{\phi_{k_1,j_1,m}'}})$ shared between Alice and Bob conditioned on measurement outcomes $k_1$ at Alice and $j_1$ at Bob, where $\rho_e$ refers to the $m$-mode system of the environment. The entropy of Alice's system is $S(\rho_A)=\log_2(d_{k_1,m})$, where the dimensionality $d_{k_1,m}$ is the number of ways $k_1$ photons can be arranged in $m$ modes, i.e., $d_{k_1,m} = {k_1+m-1 \choose k_1}$. Furthermore, since the global state $\rho_{ABE}=\ket{\phi_{k_1,j_1,m}'}{\bra{\phi_{k_1,j_1,m}'}}$ is pure, we can use the self-duality property of the von Neumann entropy to calculate the entropy of Alice and Bob's system 
\begin{align}
    S(\rho_{AB}) = S(\rho_{e}).
\end{align}
That is, the environment holds the purification of Alice and Bob's state. The entropy of the environment modes is $S(\rho_{e})=\log_2(h_{k_1,j_1,m})$, where $h_{k_1,j_1,m}$ is the number of ways $k_1-j_1$ photons can be arranged in $m$ modes, i.e., $h_{k_1,j_1,m} = {k_1-j_1+m-1 \choose k_1-j_1}$. This follows from the fact that the QND measurement is mode blind and we know for certain that $k_1-j_1$ photons were lost to the environment, but we do not know on which modes these losses occurred.

\bblack{The output state $\ket{\phi_{k_1,j_1,m}}_{ABE}$, shown in equation (3) of the main text, shows that the arrangements of photons do not depend on $\chi$ or $\eta$ and are distributed with equal weights. We confirmed this directly by computing Eve's information from the density matrix for the simplest cases. }

Therefore, using~\cref{eq:RCI}, the RCI for the renormalised state $\rho_{AB}$ is
\begin{align}
    R_{k_1,j_1,m} = S(\rho_{A})-S(\rho_{AB}) =  \log_2{\left( \frac{ d_{k_1,m} }{ h_{k_1,j_1,m} } \right)}.
\end{align}
This dimension ratio can be rewritten in a more useful form
\begin{align}
    \frac{ d_{k_1,m} }{ h_{k_1,j_1,m} } &= \frac{{k_1+m-1 \choose k_1}}{{k_1-j_1+m-1 \choose k_1-j_1}} = \frac{\frac{(k_1+m-1)!}{k_1!(m-1)!}}{\frac{(k_1-j_1+m-1)!}{(k_1-j_1)!(m-1)!}} = \frac{\frac{(k_1+m-1)!}{k_1!j_1!}}{\frac{(k_1-j_1+m-1)!}{(k_1-j_1)!j_1!}} = \frac{\frac{(k_1+m-1)!}{(k_1-j_1+m-1)!j_1!}}{\frac{k_1!}{(k_1-j_1)!j_1!}} = \frac{ {{k_1+m-1} \choose j_1} }{{ k_1 \choose j_1}} .
\end{align}
Alice is free to prepare her state offline. Hence, let us further simplify the rate expression by conditioning on Alice having already measured $k_1$ photons (i.e., fixed $k_1$ value),
\begin{align}
    S_1+F_1 &= \frac{1}{m}   \sum_{j_1=0}^{k_1} P_{k_1,j_1}^\text{Bob} \log_2{\left[ \frac{ {{k_1+m-1} \choose j_1} }{ {k_1 \choose j_1} } \right]}.
\end{align}
Note this rate does not depend on $\chi$, since this conditional probability means only Bob's measurement probability contributes to the rate $P_{k_1,j_1,m}\rightarrow  P_{k_1,j_1,m}/P^\text{Alice}_{k_1,m}=P_{k_1,j_1}^\text{Bob}$.

For small $\chi$, the most likely outcomes of Alice's QND measurement $k_1$ is a small number of photons. As $\chi$ squeezing increases, Alice will more often detect a larger number of photons. We are interested in the large energy limit $\chi\to1$, since our aim is to saturate the PLOB bound. In this limit, the squeezed states become $\ket{\chi} \propto \sum_{n=0}^\infty \ket{n}\ket{n}$, hence Alice obtains large numbers of photons $k_1\to\infty$ with unit probability. \bblack{We have proven in~\cref{app:binratioasymp} the following asymptotic relation}
\begin{align}
    \lim_{k_1\to\infty} \frac{ {{k_1+m-1} \choose j_1} }{{ k_1 \choose j_1}} = \left( 1 - \frac{j_1}{k_1}\right)^{-(m-1)}. \label{eq:binratioasym}
\end{align}
Thus, our rate expression simplifies to
\begin{align}
     \lim_{k_1\to\infty} (S_1+F_1) &=\frac{1}{m}  \lim_{k_1\to\infty} \sum_{j_1=0}^{k_1} P_{k_1,j_1}^\text{Bob} \log_2{\left[ \frac{ {{k_1+m-1} \choose j_1} }{ {k_1 \choose j_1} } \right]} \\
     &= -\frac{m-1}{m}  \lim_{k_1\to\infty} \sum_{j_1=0}^{k_1} P_{k_1,j_1}^\text{Bob} \log_2{\left(1-\frac{j_1}{k_1}\right)}. 
\end{align}
At this point, we note that $P_{k_1,j_1}^\text{Bob} = (1-\eta)^{k_1-j_1} \eta^{j_1} {k_1 \choose j_1}$ is a binomial distribution; this means that both the expected value and the most likely value Bob will measure is $j_1\approx\eta k_1$ photons. From this physical intuition, we expect only terms near $j_1\approx\eta k_1$ are important, hence the sum should simplify to $\log_2\left(1 -\eta\right)=-C(\eta)$. We can confirm this formally by firstly noting the binomial distribution moments are
\begin{align}
    \lim_{k_1\to\infty}\frac{1}{k_1^n}\langle j_1^n\rangle = \lim_{k_1\to\infty}\frac{1}{k_1^n} \sum_{j_1=0}^{k_1} P_{k_1,j_1}^\text{Bob} j_1^n = \lim_{k_1\to\infty}\frac{1}{k_1^n} \sum^n_{p=0} \left\lbrace \begin{matrix} n \\ p \end{matrix} \right\rbrace k_1(k_1-1)\cdots(k_1-p+1) \eta^p = \eta^n,
\end{align}
where we note only the $p=n$ term in the summation has the $k_1^n$ factor required to survive the $k_1\to\infty$ limit. Note that $\left\lbrace \begin{matrix} n \\ p \end{matrix} \right\rbrace$ are the Stirling numbers of the second kind (i.e., an integer independent of $k_1$), and resolves to $\left\lbrace \begin{matrix} n \\ n \end{matrix} \right\rbrace=1$ for the relevant $p=n$ case. Thus, from the Taylor series expansion of the logarithm we have
\begin{align}
     \lim_{k_1\to\infty} (S_1+F_1) &= \frac{m-1}{m}  \frac{1}{\ln{2}}\lim_{k_1\to\infty}  \sum_{j_1=0}^{k_1} P_{k_1,j_1}^\text{Bob} \sum_{n=1}^\infty \frac{j_1^n}{nk_1^n} \\
     &= \frac{m-1}{m}  \frac{1}{\ln{2}} \sum_{n=1}^\infty \frac{\eta^n}{n} \\
     &= -\frac{m-1}{m}\log_2{(1-\eta)}\\
     &= \frac{m-1}{m} C.
\end{align}
Finally, in the limit that $m\to\infty$, we have shown that the first step of our protocol is optimal since the average RCI equals the channel capacity of the pure-loss channel (the PLOB bound).

\bblack{
\subsection*{Optimality of our protocol at each round}
}

\bblack{
In this section, we show that our protocol is optimal at each round $n$.
}

\bblack{
After the first step of our protocol, the dependence on $\chi$ and on the transmissivity of the lossy channel, $\eta$, is removed from the state. That is, for some number of rails, $m$, we herald a state, $\ket{\phi}_{k,j_1,m}$, conditioned on measurement outcomes at Alice and Bob, $k_1$ and $j_1$ respectively. When $k=j_1$, the state is pure and purification is successful in the first round. The rate of this success is simply $E_{k_1=j_1} = \eta^{k_1} \log_2{{k_1+m-1}\choose k_1}$. If $k_1\neq j_1$, the state is not pure and further purification is necessary. Then, the achievable rate of further purification is given by the RCI of $\ket{\phi}_{k_1,j_1,m}$ which is $ \log_2\left [{{{k_1+m-1}\choose {k_1}} / {{k_1-j_1+m-1}\choose {k_1-j_1}}} \right]$. This quantity specifies the asymptotically-achievable rate in ebits per use of the channel of an optimal entanglement distillation protocol given infinite uses of these unsuccessful mixed states for outcomes $k_1,j_1$ where $k_1\neq j_1$. 
}

We proved in the previous section that the average rate of the successful attempt plus the optimal distillation rate equals the PLOB bound. That is, after the first round, we have the state $\ket{\phi_{{k_1},{j_1},m}}$ and we proved the average rate (over all outcomes $k_1$ and $j_1$) is given by
\begin{align}
      \lim_{\rred{\chi\to1}} \sum_{k_1=0}^{\infty} P_1 E_1 &=\frac{1}{m}  \lim_{\rred{\chi\to1}} \sum_{k_1=0}^{\infty} \sum_{j_1=0}^{k_1} (1-\chi^2)^{\frac{m}{2}} \chi^{k_1}  {{k_1+m-1} \choose k_1}   (1-\eta)^{k_1-j_1} \eta^{j_1} {k_1 \choose j_1}  \log_2{\left[ \frac{ {{k_1+m-1} \choose k_1} }{ {k_1-j_1+m-1 \choose k_1-j_1} } \right]} \label{eq:PLOB_1} \\ 
      &= \rred{\frac{1}{m}  \lim_{k_1\to\infty} \sum_{j_1=0}^{k_1}   (1-\eta)^{k_1-j_1} \eta^{j_1} {k_1 \choose j_1}  \log_2{\left[ \frac{ {{k_1+m-1} \choose k_1} }{ {k_1-j_1+m-1 \choose k_1-j_1} } \right]} \label{eq:PLOB_2}} \\ 
      &= \frac{m-1}{m} \; C ,  \label{eq:PLOB_}
\end{align}
where the sum over $j_1$ is constrained by $k_1 \geq j_1$ since Bob cannot detect more photons than Alice. If $k_1=j_1$, we have a success, otherwise if $k_1>j_1$, it fails and needs further rounds of purification. \Cref{eq:PLOB_} means we must have $m\to\infty$ in order to saturate the PLOB bound. \rred{Going from~\cref{{eq:PLOB_1}} to~\cref{{eq:PLOB_2}} shows the connection between the without pre-selection protocol and with pre-selection protocol. This equality connection is physically justified, since in the limit of large amounts of squeezed light $\chi\to\infty$, Alice should measure large amounts of photons $k_1\to\infty$ with unity probability. Going from~\cref{{eq:PLOB_2}} to~\cref{{eq:PLOB_}} was proven in the previous section.}

\bblack{
The explicit and practical entanglement distillation protocol continues in the same way as the first step, that is, Alice and Bob perform total photon number measurements on smaller subset of rails and compare results. After each successive step, the state is sometimes purified (when Alice and Bob obtain the same outcome) and sometimes it fails.
}

\bblack{
Recall that the RCI gives the achievable rate of entanglement distillation by some optimal distillation protocol. We can use it at each round, to confirm our protocol stays optimal.
}

\bblack{
Given that Alice already obtained $k_1$, and Bob already obtained $j_1$, the RCI multiplied by the success probability heralded at the $n$th round with outcomes ${k_1,j_1,\dots,k_{n-1},j_{n-1},k_n,j_n}$, for a given number of rails, $m$, is
\begin{align}
      R_n({{k_1,j_1,\dots,k_{n-1},j_{n-1}},k_n,j_n,m}) &=   \frac{ {{k_n+m-n} \choose k_n} {k_n \choose j_n} }{ {k_1+m-1 \choose k_1}{k_1 \choose j_1} }  \left[ \prod_{s=1}^{n-1} {{k_s-k_{s+1}} \choose {j_s-j_{s+1}}} \right]  \log_2{\left[ \frac{ {{k_n+m-n} \choose k_n} }{ {k_n-j_n+m-n \choose k_n-j_n} } \right]},
\end{align}
where we have the constraints: 
\begin{align}
    &k_1 \geq k_2 \geq k_3 \geq \dots \geq k_n,\\
    &j_1 \geq j_2 \geq j_3 \geq \dots \geq j_n,\\
    &k_s-k_{s+1} \geq j_s-j_{s+1} \; \forall s ,\\
    &k_s > j_s \; \forall s \neq n, \\
    &k_n=j_n.
\end{align}
Intuitively, the first (second) constraint $k_{s} \geq k_{s+1}$ ($j_{s} \geq j_{s+1}$) just says Alice (Bob) can't measure more photons in the next round $s+1$ compared to the current round $s$. The third constraint $k_s-k_{s+1} \geq j_s-j_{s+1}$ simply says the amount of photons in the left-over rail (i.e. the rail we are throwing away) can't have more photons on Bob's side $j_s-j_{s+1}$ compared to Alice's side $k_s-k_{s+1}$. The fourth constraint $k_s>j_s$ is the condition where we continue the distillation process. If $k_n=j_n$ then the RCI is equal to the von Neumann entropy since the state is pure. This counts as a successes, otherwise further purification is required.
}

\bblack{
Then, the average RCI of the $n$th round, $R_n$, over all outcomes of that round only, $k_n$ and $j_n$, consists of pure terms (von Neumann entropies) and mixed terms (RCI), and should be equal to the RCI of the previous round in the limit of large $m$. That is, we must have that no entanglement is lost at each round. So we must have $\sum_{k_n,j_n} P_n R_n = P_{n-1} E_{n-1}$ in the limit of large $m$, i.e., we must have:
\begin{align}
      \sum_{k_n} \sum_{j_n}  \frac{ {{k_n+m-n} \choose k_n} {k_n \choose j_n} }{ {k_1+m-1 \choose k_1}{k_1 \choose j_1} } \nonumber & \left[ \prod_{s=1}^{n-1} {{k_s-k_{s+1}} \choose {j_s-j_{s+1}}} \right]  \log_2{\left[ \frac{ {{k_n+m-n} \choose k_n} }{ {k_n-j_n+m-n \choose k_n-j_n} } \right]}     \\&=  \frac{ {{k_{n-1}+m-n+1} \choose k_{n-1}} {k_{n-1} \choose j_{n-1}} }{ {k_1+m-1 \choose k_1}{k_1 \choose j_1} }  \left[ \prod_{s=1}^{n-2} {{k_s-k_{s+1}} \choose {j_s-j_{s+1}}} \right]  \log_2{\left[ \frac{ {{k_{n-1}+m-n+1} \choose k_{n-1}} }{ {k_{n-1}-j_{n-1}+m-n+1 \choose k_{n-1}-j_{n-1}} } \right]},\label{eq:??}
\end{align}
where we have the usual constraints on all the outcomes. We have numerically verified that this is true. Numerically we have verified that the left hand side approaches the right hand side of~\cref{eq:??} multiplied by the factor $\frac{m-1}{m}$, i.e., $\sum_{k_n} \sum_{j_n}  \frac{ {{k_n+m-n} \choose k_n} {k_n \choose j_n} }{ {k_1+m-1 \choose k_1}{k_1 \choose j_1} } \nonumber  \left[ \prod_{s=1}^{n-1} {{k_s-k_{s+1}} \choose {j_s-j_{s+1}}} \right]  \log_2{\left[ \frac{ {{k_n+m-n} \choose k_n} }{ {k_n-j_n+m-n \choose k_n-j_n} } \right]}     = \frac{m-1}{m}  \frac{ {{k_{n-1}+m-n+1} \choose k_{n-1}} {k_{n-1} \choose j_{n-1}} }{ {k_1+m-1 \choose k_1}{k_1 \choose j_1} }  \left[ \prod_{s=1}^{n-2} {{k_s-k_{s+1}} \choose {j_s-j_{s+1}}} \right]  \log_2{\left[ \frac{ {{k_{n-1}+m-n+1} \choose k_{n-1}} }{ {k_{n-1}-j_{n-1}+m-n+1 \choose k_{n-1}-j_{n-1}} } \right]}$. The factor $\frac{m-1}{m}$ comes from the combinatorics of counting photons at Alice's and Bob's side.
}

\bblack{
We cancel the common factors and explicitly write the limits on $k_n,j_n$:
\begin{align}
      \sum_{k_n=1}^{k_{n-1}} \sum_{j_n={\text{max}(1,j_{n-1}-k_{n-1}+k_n})}^{\text{min}(k_n,j_{n-1})}  { {{k_n+m-n} \choose k_n} {k_n \choose j_n} } \nonumber &  {{k_{n-1}-k_{n}} \choose {j_{n-1}-j_{n}}}  \log_2{\left[ \frac{ {{k_n+m-n} \choose k_n} }{ {k_n-j_n+m-n \choose k_n-j_n} } \right]}     \\&=   { {{k_{n-1}+m-n+1} \choose k_{n-1}} {k_{n-1} \choose j_{n-1}} }   \log_2{\left[ \frac{ {{k_{n-1}+m-n+1} \choose k_{n-1}} }{ {k_{n-1}-j_{n-1}+m-n+1 \choose k_{n-1}-j_{n-1}} } \right]}.\label{eq:??}
\end{align}
We may consider the asymptotic behaviour for a large number of rails.
}

\bblack{
Note that for large $m$ the most important term is when Alice measures $k_n=k_{n-1}$, which means Bob measures $j_n=j_{n-1}$. This term is the following
\begin{align}
    &\binom{k_{n-1}+m-n}{k_{n-1}} \binom{k_{n-1}}{j_{n-1}} \log_2 \left[ \frac{ \binom{k_{n-1}+m-n}{k_{n-1}} }{ \binom{k_{n-1}-j_{n-1}+m-n}{k_{n-1}-j_{n-1}} } \right] \nonumber \\
    &= \frac{m-n+1}{k_{n-1}+m-n+1}\binom{k_{n-1}+m-n+1}{k_{n-1}} \binom{k_{n-1}}{j_{n-1}} \log_2 \left[ \frac{ \frac{m-n+1}{k_{n-1}+m-n+1} }{ \frac{m-n+1}{k_{n-1}-j_{n-1}+m-n+1} } \frac{ \binom{k_{n-1}+m-n+1}{k_{n-1}} }{ \binom{k_{n-1}-j_{n-1}+m-n+1}{k_{n-1}-j_{n-1}} } \right] \\ 
    &= \frac{m-n+1}{k_{n-1}+m-n+1}\binom{k_{n-1}+m-n+1}{k_{n-1}} \binom{k_{n-1}}{j_{n-1}} \left( \log_2 \left[ \frac{ \binom{k_{n-1}+m-n+1}{k_{n-1}} }{ \binom{k_{n-1}-j_{n-1}+m-n+1}{k_{n-1}-j_{n-1}} } \right] +  \log_2 \left[ \frac{k_{n-1}-j_{n-1}+m-n+1}{k_{n-1}+m-n+1} \right] \right) 
\end{align}
where we used the following identity $\binom{k+m}{k} = \frac{(k+m)!}{k!m!} = \frac{m+1}{k+m+1} \frac{(k+m+1)!}{k!(m+1)!} = \frac{m+1}{k+m+1} \binom{k+m+1}{k}$. We can then consider the ratio between this final term and the RHS of Eq.~\cref{eq:??} as follows
\begin{align}
    &\frac{ \frac{m-n+1}{k_{n-1}+m-n+1} \binom{k_{n-1}+m-n+1}{k_{n-1}} \binom{k_{n-1}}{j_{n-1}} \left( \log_2 \left[ \frac{ \binom{k_{n-1}+m-n+1}{k_{n-1}} }{ \binom{k_{n-1}-j_{n-1}+m-n+1}{k_{n-1}-j_{n-1}} } \right] +  \log_2 \left[ \frac{k_{n-1}-j_{n-1}+m-n+1}{k_{n-1}+m-n+1} \right] \right) }{ \binom{k_{n-1}+m-n+1}{k_{n-1}} \binom{k_{n-1}}{j_{n-1}} \log_2 \left[ \frac{ \binom{k_{n-1}+m-n+1}{k_{n-1}} }{ \binom{k_{n-1}-j_{n-1}+m-n+1}{k_{n-1}-j_{n-1}} } \right]  } \nonumber \\ 
    &= \frac{m-n+1}{k_{n-1}+m-n+1}  \left( 1 + \frac{ \log_2 \left[ \frac{k_{n-1}-j_{n-1}+m-n+1}{k_{n-1}+m-n+1} \right] }{ \log_2 \left[ \frac{ \binom{k_{n-1}+m-n+1}{k_{n-1}} }{ \binom{k_{n-1}-j_{n-1}+m-n+1}{k_{n-1}-j_{n-1}} } \right] }\right) \xrightarrow{m \to \infty} 1
\end{align}
This ratio approaches one in the limit as $m\to\infty$ (note that we also require $k_{n-1}/m\rightarrow 0$), because we can show that we have 
\begin{align}
    \frac{m-n+1}{k_{n-1}+m-n+1}  &= \frac{1-n/m+1/m}{k_{n-1}/m+1-n/m+1/m}  \xrightarrow{m \to \infty} 1 \\
    \frac{k_{n-1}-j_{n-1}+m-n+1}{k_{n-1}+m-n+1} &= \frac{k_{n-1}/m-j_{n-1}/m+1-n/m+1/m}{k_{n-1}/m+1-n/m+1/m} \xrightarrow{m \to \infty} 1 \\ 
     \frac{ \binom{k_{n-1}+m-n+1}{k_{n-1}} }{ \binom{k_{n-1}-j_{n-1}+m-n+1}{k_{n-1}-j_{n-1}} } &=  \frac{ \binom{k_{n-1}+m-n+1}{j_{n-1}} }{ \binom{k_{n-1}}{j_{n-1}} } \xrightarrow{m \to \infty} \infty \\ 
     &\Rightarrow \frac{ \log_2 \left[ \frac{k_{n-1}-j_{n-1}+m-n+1}{k_{n-1}+m-n+1} \right] }{ \log_2 \left[ \frac{ \binom{k_{n-1}+m-n+1}{k_{n-1}} }{ \binom{k_{n-1}-j_{n-1}+m-n+1}{k_{n-1}-j_{n-1}} } \right] } \xrightarrow{m \to \infty} 0 
\end{align}
Thus we have shown that no entanglement is lost at the $n$th round, in the limit that we have a large amount of rails $m$. 
}

\bblack{This optimally can be motivated by considering the environment. We showed that for $m\to\infty$, we have that no RCI is lost after every round:
\begin{align}
   S_n + F_n &= F_{n-1}.
\end{align}
}

\bblack{
When the protocol succeeds, it decouples the environment from Alice and Bob, and when it fails it does not (but may succeed in a later round). Alice and Bob's measurement tells them how \rred{many} photons were lost to the environment. If we consider the environment we do not need to average over the particular outcomes at Alice and Bob so we can avoid having to compute the complicated sum over many outcomes. If Alice and Bob forget the particular outcomes but \rred{keep} the environment outcome (whether the photon number at Alice and Bob was the same or different) then effectively they have measured the environment with either a success or \rred{failure}: Our quantum measurement is defined by the set of operators $\{M_S, M_F\}$ acting on the environment, where $M_S = \ketbra{\bm{0}}{\bm{0}}$ and $M_F= \Id-\ketbra{\bm{0}}{\bm{0}}$, satisfying $M_S+M_F=\Id$.
So, after the first round, the global successful and failed density matrices shared between Alice, Bob, and the environment are:
\begin{align}
    \rho_{S_1} &= \ketbra{\bm{0}_m}{\bm{0}_m} \rho \ketbra{\bm{0}_m}{\bm{0}_m}  = \text{decoupled from the environment}, \label{eq:S1}\\
    \rho_{F_1} &= (\Id_m-\ketbra{\bm{0}_m}{\bm{0}_m}) \rho (\Id_m-\ketbra{\bm{0}_m}{\bm{0}_m})  = \text{not decoupled from the environment},\label{eq:F1}
\end{align}
where $\ket{\bm{0}_m}=\ket{0000\dots}$ (i.e., $m$ vacuum modes), $\Id_m$ is the identity on $m$ modes, and $\rho$ is the global system of the $m$ initial copies of lossy TSMV state with squeezing $\chi$ and transmissivity $\eta$.~\Cref{eq:S1} projects the system onto vacuum of the environment, purifying completely Alice and Bob's state, while~\cref{eq:F1} does not. The density matrix transforms as un\rred{n}ormalised density matrix $\rho\to M_i \rho M_i^\dagger$, where $P_i=\Tr( \rho M_i^\dagger M_i)$ is the probability.
We have (for $m\to\infty,\chi\to1$):
\begin{align}
    R(\rho_{S_1}) + R(\rho_{F_1}) &= R(\rho) = C,\label{eq:Env}
\end{align}
which agrees with what we found earlier, that the first round is optimal and loses no rate as $m\to\infty$.
}

\bblack{
Before round two, we have the failed density matrix from round one: $\rho_{F_1}$. After round two, it either succeeds or fails (while the $m$th mode is left alone and is completely decoupled from the system anyway), and we have:
\begin{align}
    \rho_{S_2} &= (\ketbra{\bm{0}_{m-1}}{\bm{0}_{m-1}} \otimes \Id_1 ) \rho_{F_1} (\ketbra{\bm{0}_{m-1}}{\bm{0}_{m-1}} \otimes \Id_1 )  = \text{decoupled from the environment},\\
    \rho_{F_2} &= [(\Id_{m-1}-\ketbra{\bm{0}_{m-1}}{\bm{0}_{m-1}}) \otimes \Id_1] \rho_{F_1} [(\Id_{m-1}-\ketbra{\bm{0}_{m-1}}{\bm{0}_{m-1}}) \otimes \Id_1] = \text{not decoupled from the environment}.
\end{align}
}

\bblack{
So we have
\begin{align}
    R(\rho_{S_2}) + R(\rho_{F_2}) &=  R(\rho_{F_1}),
\end{align}
which can be shown to be true if we write $\rho_{F_1}$ in terms of its eigenvalues.
}

\bblack{
After the $n$th round, we have:
\begin{align}
    \rho_{S_n} &= (\ketbra{\bm{0}_{m+n-1}}{\bm{0}_{m+n-1}} \otimes \Id_{n-1} ) \rho_{F_n} (\ketbra{\bm{0}_{m+n-1}}{\bm{0}_{m+n-1}} \otimes \Id_{n-1} ),\\
    \rho_{F_n} &= [(\Id_{m+n-1}-\ketbra{\bm{0}_{m+n-1}}{\bm{0}_{m+n-1}}) \otimes \Id_{n-1}] \rho_{F_n} [(\Id_{m+n-1}-\ketbra{\bm{0}_{m+n-1}}{\bm{0}_{m+n-1}}) \otimes \Id_{n-1}],
\end{align}
and we have
\begin{align}
    R(\rho_{S_n}) + R(\rho_{F_n}) &= R(\rho_{F_{n-1}}).
\end{align}
}

\bblack{
\subsection*{Proof that the highest rate of our iterative purification protocol achieves the PLOB bound}
}

\bblack{There is a straight-forward recurrence relation in determining the average RCI for success and failure at the $n$th round of purification
\begin{align}
    S_1 + F_1  &=   \frac{m-1}{m} C, \\
   S_{n} + F_{n} &\simeq \frac{m-1}{m} F_{n-1}, \label{eq:F_n_iterative}
\end{align}
where $n\in[2,m]$ and $\simeq$ means asymptotic equivalence given a large number of rails $m$. Intuitively, this relationship simply says that about $1/m$ fraction of the left-over rate is lost at each round. We want to show that the total rate of our iteration protocol, which is the sum of all the success rates 
\begin{align}
    E_{\text{iteration},m} &\equiv \sum_{n=1}^m S_n \simeq  \frac{m-1}{m} C - \sum_{n=1}^{m}  \frac{F_n}{m}, 
\end{align}
will add up to the capacity given infinite resources $E_\text{iteration} \equiv \lim_{m\rightarrow\infty} E_{\text{iteration},m} = C$. This is not a trivial question to answer, as the subtraction of $m$ of these $F_n/m$ terms has the \textit{possibility} of significantly reducing the rate of our protocol, even in $m\rightarrow\infty$ limit. However, we show here that this is not the case, and hence our protocol can achieve the PLOB bound capacity $C$.} 

\bblack{
Now, starting from the recurrence expression in~\cref{eq:F_n_iterative}, we can derive the following closed-form expression for the failure rate at the $n$th round
\begin{align}
    F_{n} &\simeq \left( \frac{m-1}{m} \right)^n C - \sum_{i=1}^n \left( \frac{m-1}{m} \right)^{n-i} S_i. \label{eq:F_n_expanded}
\end{align}
The sum of these rates can be simplified as follows 
\begin{align}
    \sum_{n=1}^{m} \frac{F_n}{m} &\simeq  \sum_{n=1}^m \left( \frac{m-1}{m} \right)^n \frac{C}{m} - \sum_{n=1}^m \sum_{i=1}^n  \left( \frac{m-1}{m} \right)^{n-i} \frac{S_i}{m} \\
    &\simeq \frac{C}{m} \sum_{n=1}^m \left( \frac{m-1}{m} \right)^n - \sum_{i=1}^m \frac{S_i}{m}  \sum_{n=i}^m \left( \frac{m-1}{m} \right)^{n-i}  \\
    &\simeq \frac{m-1}{m} C \left[ 1 - \left( \frac{m-1}{m} \right)^m \right] - \sum_{i=1}^m S_i \left[ 1- \left( \frac{m-1}{m} \right)^{1+m-i} \right]. 
\end{align}
In the asymptotic limit, this reduces down to 
\begin{align}
    \lim_{m\to\infty} \sum_{n=1}^{m} \frac{F_n}{m} &= C \left( 1-\frac{1}{e} \right) -   \sum_{i=1}^\infty S_i \left( 1-\frac{1}{e} \right) = C  \left( 1-\frac{1}{e} \right) - E_\text{iteration} \left( 1-\frac{1}{e} \right). 
\end{align}
Hence, we can show that the total rate of our iteration protocol gives 
\begin{align}
    E_\text{iteration} &= C - \lim_{m\to\infty} \sum_{n=1}^m \frac{F_n}{m} = \frac{C}{e} + E_\text{iteration} - \frac{E_\text{iteration}}{e}, \\
    \therefore E_\text{iteration} &= C.
\end{align}
Thus we have proven that our iteration protocol can achieve the PLOB bound. 
}

\bblack{
\subsection*{Entropy argument for achieving the PLOB bound}
}

\bblack{In this section, we give a simpler argument that our protocol is optimal based on von Neumann entropies.}

\bblack{
The RCI gives the achievable rate of entanglement distillation of an optimal protocol:
\begin{align}
    E_d &\geq R = S(\rho_A)-S(\rho_e).
\end{align}
For the lossy channel we have
\begin{align}
    S(\rho_A)-S(\rho_e) = C.
\end{align}
Our goal is to specify a protocol which achieves $C$ (PLOB).
}

\bblack{
Imagine that Alice and Bob perform the following purification protocol. Alice shares entanglement with Bob in the form of TMSV states across a lossy channel then they both locally count total photon number on multiple modes and compare outcomes. Next, they keep the times they obtained the same outcome but forget what the particular outcomes were. This decouples the environment and Alice and Bob share mixed states. This procedure is equivalent to Eve performing a QND measurement on the environment and heralding states shared between Alice and Bob corresponding to no lost photons in some subset of rails. In the $n$th iterative round, all parties, Alice, Bob, and Eve, learn if any photons were lost to the environment in the first $m-n+1$ rails. We only keep the times no photons were lost. We must take $m\to\infty$ because the RCI is asymptotic in $m$ and it is not optimal to operate and perform measurements on less than infinite copies of states. Then we must at least achieve the RCI in the limit as $m\to\infty$ since we argue elsewhere that no entanglement is lost after each round, and we have the rate of our protocol
\begin{align}
    E_\text{iteration} &\geq \sum_{n=1}^{m-1} P_n (S(\rho_{n_A})-S(\rho_{n_{AB}})) = R = -\log_2{(1-\eta)},
\end{align}
where $P_n$ is the probability of success of the $n$th round, and $\rho_{n_{AB}}$ is the mixed state shared between Alice and Bob which is decoupled from Eve. $\rho_{n_A}=\Tr_B{\rho_{n_{AB}}}$ is Alice's reduced state.
}

\bblack{
Finally, is there a simple protocol which purifies the states $\rho_{n_{AB}}$ at a rate at least as high as that given by the RCI? Counting photons purifies at this rate as $m\to\infty$ since the states heralded by Alice and Bob's outcome, $k_n$, at the $n$th round are pure with entanglement $S(\rho_{k_{n_{AB}}}) = \log_2{ {{k_n+m-n} \choose k_n} } $, and we have that $S(\rho_{k_{n_{AB}}})=0$ for all $n$. Thus, for $m\to\infty$ and $\chi\to1$, the achievable rate of our protocol is
\begin{align}
    E_\text{iteration} &= \sum_{n=1}^{m-1} \sum_{k_n} P_{k_n} \log_2{ {{k_n+m-n} \choose k_n} } = R = -\log_2{(1-\eta)} = C,
\end{align}
so we purify completely at the PLOB rate. The equality follows from the fact that the states are pure so no more entanglement can be purified from them. This makes sense since the PLOB is an upper bound. We do not need to know the values of $P_{k_n}$ to achieve our goal. Obtaining a simple expression for $P_{k_n}$ seems intractable because the sum over all outcomes is complicated.
}

\bblack{
\subsection*{Numerical results for the exact rate of our iterative purification protocol (finite $m$)}
}

\bblack{
The rate of our purification protocol (in ebits per use) is maximised if Alice performs her first measurement offline, where she obtains outcome $k_1$. For finite $m$, there is an optimal choice of $k_1$.
}

\begin{figure}
\centering
\includegraphics[width=0.5\linewidth]{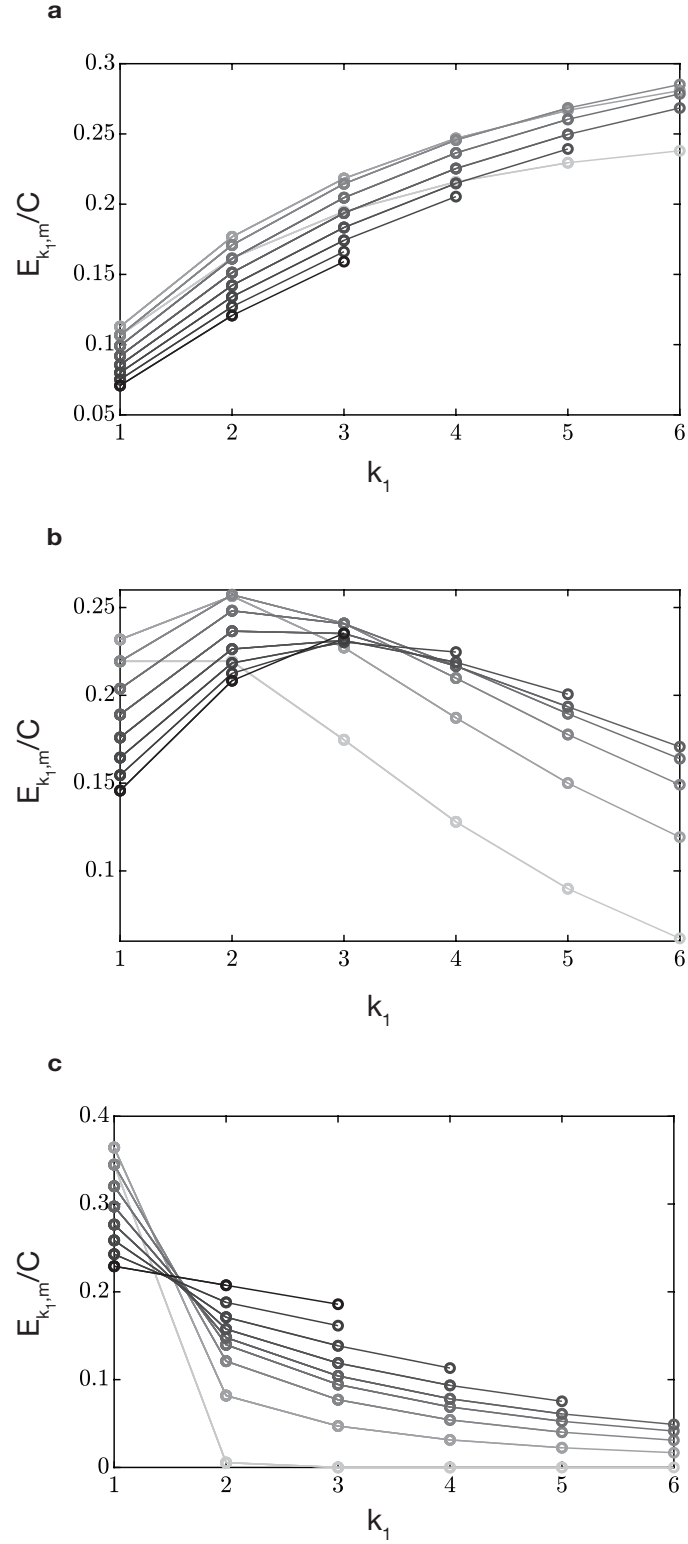}
\caption{Ratio of our finite $m$ rate, $E_{k_1,m}$ with the two-way assisted quantum capacity, $C$, (PLOB)~\cite{SUP_Pirandola_2017} as a function of $k_1$ at Alice at a fixed distance of \textbf{a} 1 km, \textbf{b} 10 km, and \textbf{c} 100 km. The number of rails ranges from $m=2$ (light gray) to $m=9$ (black). The figure shows that for finite $m$ there is an optimal $k_1$ at a particular distance. Unfortunately, we cannot simulate large numbers of rails because we have to sum over many measurement outcomes. Analytically, we prove that $E_{k_1,m}/C$ $\to 1 $ as $m\to\infty$ and $k_1\to \infty$.\label{fig:finite_m_rate}}
\end{figure}

\bblack{
We numerically compute the rate for finite $m$ (and include up to $n=m-1$ rounds in the sum shown by equation (4) of the main text). We plot it in~\cref{fig:finite_m_rate}. Iteration becomes useful for $k_1>1$. Even for small $k_1$, the rates are quite close to the PLOB bound.
}

\bigskip

\section{Entanglement swapping}\label{app:ent_swapping}

In this Supplementary Note, we discuss how our repeaterless purification scheme can be used to distribute entanglement between nodes of a larger quantum network.

\begin{figure}
\centering
\includegraphics[width=0.7\linewidth]{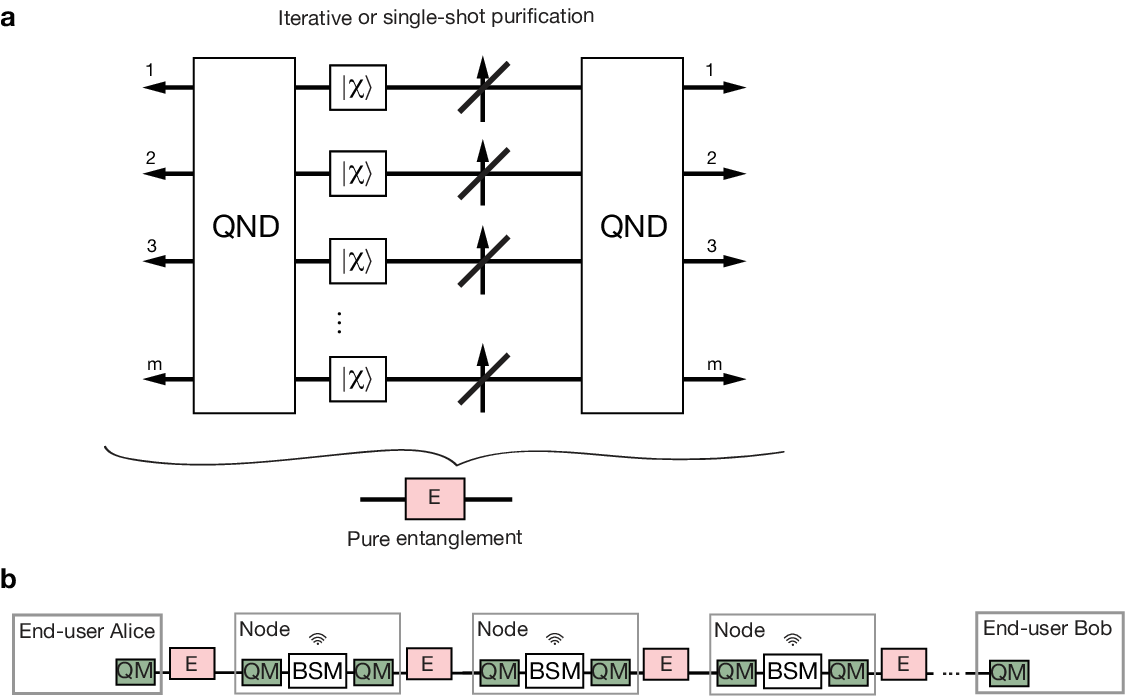}
\caption{\textbf{Our purification for large quantum networks.} \textbf{a} Our purification technique operating between neighbouring nodes in a network. It can be iterative or single-shot (non-iterative). \textbf{b} Linear quantum chain as an example of a simple quantum network. First, entanglement is purified between all neighbouring nodes of the network (pink) and the pure entanglement is held in quantum memories (QM, green). Then, Bell-state measurements (BSM) and classical communication perform quantum teleportation (i.e., entanglement swapping) so that eventually the end parties share pure entanglement. The Bell-state measurements can be DV or CV.}
\label{fig:purification_repeater_protocol}
\end{figure}

Consider a linear quantum network for simplicity, as shown in~\cref{fig:purification_repeater_protocol}, where $N-1$ untrusted quantum repeaters divide the total distance between the end users into $N$ links. Neighbouring nodes perform purification and store the purified states in ideal quantum memories. The quantum memories allow the probabilistic purification schemes of each link to succeed independently from each other. Once purification is complete, pure entanglement is shared between all neighbouring nodes of the network.

To distribute entanglement between the end users the nodes simply perform entanglement swapping via Bell-state measurements. \bblack{The swapping happens in parallel assuming many copies of the protocol are available. For the iterative scheme, the dimension of the states will most likely be different so the nodes should link up the states with the same dimension. This is always possible if they have access to many copies held in quantum memories. Otherwise, if the states have different dimensions then entanglement swapping is limited to the smallest dimension. }

\bblack{Alternatively, the nodes could transfer all pure entanglement onto qubits before the swapping. Clearly, single-shot purification is more practical since all states have dimension $d_{k,m}$ and $d_{k,m}$-dimensional swapping is done on those states.}

The entanglement swapping can be DV or CV. For DV, the heralded pure maximally-entangled output states have dimension \bblack{$d_{k_n,m} = {{k_n+m-n} \choose {k_n}}$}, then the DV Bell-state measurement at the node projects the pure input state onto one of the known \bblack{$d_{k_n,m}^2$} Bell states with success probability \bblack{$1/d_{k_n,m}^2$}. Hence, with a success probability of one, a maximally entangled state is heralded between non-neighbouring nodes. The nodes communicate classically which Bell-state measurement outcome was obtained and based on this information the non-neighbouring nodes can perform the required correction. Eventually, the end users share pure entanglement. Crucially, the entanglement swapping is deterministic, therefore, assuming ideal quantum memories and perfect entanglement-swapping operations, the highest rates of our purification scheme saturate the ultimate end-to-end rates of lossy quantum communications. 

For CV, first, assume that non-maximally pure entangled states are shared between all neighbouring nodes of the network which are copies of the approximately-Gaussian truncated TMSV state
\begin{align}
    \ket{\chi_{k,m}} &=  \sqrt{\frac{\chi^2-1}{\chi^{2d_{k,m}}-1}} \sum_{\mu=0}^{d_{k,m}-1} \chi^\mu \ket{\mu}_A \ket{\mu}_B.
\end{align}
Our single-shot purification scheme (quantum error detection) can be used to distribute $\ket{\chi_{k,m}}$ between any two neighbouring nodes of a lossy quantum network for a given $k$ and $m$.

Next, require that $\ket{\chi_{k,m}}$ is sufficiently Gaussian by choosing $d$ large enough for a given $\chi$. This allows us to use the covariance-matrix formalism. Then, we consider a deterministic CV Bell-state measurement in the relay consisting of dual homodyne, i.e., mixing the two incoming modes on a 50:50 beamsplitter and performing orthogonal quadrature measurements. The nodes communicate the outcomes of the dual-homodyne measurements so that the required displacement-correction operations can be performed. Since the state is sufficiently Gaussian, the covariance matrix is all that is required to calculate the rate of distilled entanglement or estimate the secret key rate. The covariance matrix does not depend on the particular outcome of the dual-homodyne measurement, and it can be calculated easily via the symplectic transformations for the dual-homodyne measurements as follows.

Assuming the dimension, $d_{k,m}$, is sufficiently large, then the state $\ket{\chi_{k,m}}$ is approximately Gaussian and the state is fully characterised by the following covariance matrix~\cite{SUP_Weedbrook_2012}:
\begin{align}
   V &= \begin{bmatrix}
        \nu & 0 & \sqrt{\nu^2-1} & 0\\
        0 & \nu & 0 & -\sqrt{\nu^2-1}\\
        \sqrt{\nu^2-1} & 0 & \nu & 0\\
        0 & -\sqrt{\nu^2-1} & 0 & \nu\\
    \end{bmatrix}.\label{eq:CM_chi}
\end{align}
where $\nu=\cosh{2r}$ with $\chi=\tanh{r} \in [0,1]$.

We start with two copies of this state, i.e., two links connected by a single repeater. Mode 1 is entangled with mode 2 and mode 3 is entangled with mode 4. Dual-homodyne consists of a 50:50 beamsplitter followed by orthogonal quadrature measurements. Let modes 1 and 4 be the two modes input to the quantum repeater and modes 2 and 3 be non-neighbouring nodes in the network. Modes 2 and 3 will remain and be entangled after the entanglement swapping operation. Modes 1 and 4 are input on a 50:50 beamsplitter transforms the four-mode covariance matrix like
\begin{align}
   \begin{bmatrix} V & \\ & V \end{bmatrix} &\to S_B \begin{bmatrix} V & \\ & V \end{bmatrix} S_B^T,
\end{align}
where
\begin{align}
   S_B &= \begin{bmatrix}
        \frac{1}{\sqrt{2}} & 0 & 0 & 0 & 0& 0 & \frac{1}{\sqrt{2}} & 0\\
        0 & \frac{1}{\sqrt{2}} & 0 & 0 & 0 & 0& 0 & \frac{1}{\sqrt{2}}\\
         0 & 0 & 1 & 0 & 0 & 0 & 0& 0\\
         0 & 0 & 0 & 1 & 0 & 0 & 0& 0\\
         0 & 0 & 0 & 0 & 1 & 0 & 0& 0\\
         0 & 0 & 0 & 0 & 0 & 1 & 0& 0\\
        -\frac{1}{\sqrt{2}} & 0 & 0 & 0& 0 & 0  & \frac{1}{\sqrt{2}} & 0\\
        0 & -\frac{1}{\sqrt{2}} & 0 & 0& 0 & 0 & 0 & \frac{1}{\sqrt{2}}\\
    \end{bmatrix}.
\end{align}

After the beamsplitter, the four-mode quantum system becomes
\begin{align}
   &=  \begin{bmatrix}
                              \nu   &                                0   &     \frac{ \sqrt{2(\nu^2-1)}}{2}   &                                0   &    \frac{ \sqrt{2(\nu^2-1)}}{2}   &                                0   &                                0   &                                0\\
                            0   &                               \nu   &                                0   &    -\frac{ \sqrt{2(\nu^2-1)}}{2}   &                               0   &    -\frac{ \sqrt{2(\nu^2-1)}}{2}   &                                0   &                                0\\
 \frac{ \sqrt{2(\nu^2-1)}}{2}   &                                0   &                               \nu   &                                0   &                               0   &                                0   &    -\frac{ \sqrt{2(\nu^2-1)}}{2}   &                                0\\
                            0   &    -\frac{ \sqrt{2(\nu^2-1)}}{2}   &                                0   &                               \nu   &                               0   &                                0   &                                0   &     \frac{ \sqrt{2(\nu^2-1)}}{2}\\
 \frac{ \sqrt{2(\nu^2-1)}}{2}   &                                0   &                                0   &                                0   &                              \nu   &                                0   &     \frac{ \sqrt{2(\nu^2-1)}}{2}   &                                0\\
                            0   &    -\frac{ \sqrt{2(\nu^2-1)}}{2}   &                                0   &                                0   &                               0   &                               \nu   &                                0   &    -\frac{ \sqrt{2(\nu^2-1)}}{2}\\
                            0   &                                0   &    -\frac{ \sqrt{2(\nu^2-1)}}{2}   &                                0   &    \frac{ \sqrt{2(\nu^2-1)}}{2}   &                                0   &                               \nu   &                                0\\
                            0   &                                0   &                                0   &     \frac{ \sqrt{2(\nu^2-1)}}{2}   &                               0   &    -\frac{ \sqrt{2(\nu^2-1)}}{2}   &                                0   &                               \nu\\
                  \end{bmatrix}.
\end{align}

For a matrix written in block form:
\begin{align}
    \Gamma = \begin{bmatrix} \Gamma_X & \sigma \\ \sigma^T & \Gamma_Y \end{bmatrix},
\end{align}
measuring a quadrature of subsystem $\Gamma_Y$ via homodyne detection, where $\Gamma_Y$ is a $2\times2$ real matrix, transforms the covariance matrix of subsystem $\Gamma_X$ as follows~\cite{SUP_Weedbrook_2012}:
\begin{align}
    \Gamma &\to \Gamma_X-\sigma(\Pi \;  \Gamma_Y \; \Pi)^{-1} \sigma^T,\label{eq:hom_CM}
\end{align}
where $\Pi = \text{diag}(1,0)$ for $\hat{q}$ quadrature and $\Pi = \text{diag}(0,1)$ for $\hat{p}$ quadrature, and $(\Pi \;  \Gamma_Y \; \Pi)^{-1}$ is a pseudoinverse. Note that the output covariance matrix does not depend on the measurement outcome.

For our four-mode system, we homodyne modes 1 and 4 in the repeater, one in $\hat{q}$ quadrature and one in $\hat{p}$ quadrature. We use~\cref{eq:hom_CM} twice since we perform two measurements. Performing homodyne in $\hat{q}$ quadrature of mode 1 and $\hat{p}$ quadrature of mode 4, the resulting covariance matrix is
\begin{align}
 \begin{bmatrix}
      \frac{\nu^2+1}{2\nu}  &                     0  &     -\frac{\nu^2-1}{2\nu}  &                     0\\
                  0  &     \frac{\nu^2+1}{2\nu}  &                      0  &     \frac{\nu^2-1}{2\nu}\\
 -\frac{\nu^2-1}{2\nu}  &                     0  &      \frac{\nu^2+1}{2\nu}  &                     0\\
                  0  &     \frac{\nu^2-1}{2\nu}  &                      0  &     \frac{\nu^2+1}{2\nu}\\
                  \end{bmatrix},
\end{align}
so we can see that the output state is a TMSV state with reduced squeezing, quantified by the reduced noise variance $\tilde{\nu} = \frac{\nu^2+1}{2\nu}$. That is, where previously two non-neighbouring nodes were not entangled, now they share deterministically a pure TMSV state with reduced squeezing. We keep doing this over all repeater nodes such that eventually the end users share deterministically a TMSV state with covariance matrix of the same form with a new noise variance which depends on the number of repeater nodes and initial squeezing. This covariance matrix shared between the end users allows us to estimate the asymptotic secret key rate.

A lower bound on the asymptotic secret key rate in the case of reverse reconciliation for collective attacks is given by the Devetak-Winter rate~\cite{SUP_devetak2005distillation}
\begin{equation}
K_{DW} = \beta H_{AB} - \chi_{EB},
\end{equation}
where $H_{AB}$ is the classical (Shannon) mutual information between the trusted end users, which we now call end-user Alice and end-user Bob, $\chi_{EB}$ is the Holevo quantity, the maximal quantum mutual information between Eve and end-user Bob (the reference side of the information reconciliation), and $\beta$ is the reconciliation efficiency. Conveniently, for our purification protocol the end-users are identical so can be swapped (i.e., there is no difference between reverse reconciliation and direct reconciliation).

The upper bound on the information extractable by Eve is given by the Holevo quantity~\cite{SUP_holevo1973bounds}
\begin{equation}
\chi_{EB} = S(\rho_E) - S(\rho_{E|b}),\label{eq:HOLEVO}
\end{equation}
where $S(\rho_E)$ is the von Neumann entropy of Eve's state, and $S(\rho_{E|b})$ is the von Neumann entropy of Eve's state conditioned on end-user Bob's homodyne measurement. 

The Holevo information is obtained by allowing Eve access to the purification of the state shared between the end-users $\rho_{AB}$. The global state $\rho_{ABE}$ is pure and we can use the self-duality property of the von Neumann entropy to write $S(\rho_E)=S(\rho_{AB})$ and $S(\rho_{E|b})=S(\rho_{A|b})$, where $\rho_{A|b}$ is end-user Alice's mode conditioned on a measurement on end-user Bob's mode. 

For Gaussian states, $\chi_{EB}$ may be calculated from the symplectic eigenvalues of the covariance matrices of $\rho_{AB}$ and $\rho_{A|b}$ in the entanglement-based CV-QKD protocol. Thus, Eve's information is given by
\begin{equation}
\chi_{EB} = S(\rho_{AB}) - S(\rho_{A|b}),\label{eq:SecretKeyRate}
\end{equation}
i.e., the Holevo quantity is given in terms of von Neumann entropies which can be calculated using the symplectic eigenvalues $\nu_k$ of the covariance matrix of the state~\cite{SUP_PhysRevA.59.1820}, i.e., $V_{AB}$ and $V_{A|b}$, via the relation $S(\rho)=\sum_{k=1}^N \frac{\nu_k+1}{2}\log_2 \frac{\nu_k+1}{2} - \frac{\nu_k-1}{2}\log_2 \frac{\nu_k-1}{2},$ where $N$ is the number of modes.

\bblack{Entanglement is lost when distributing approximate TMSV states, captured by the covariance matrix transformation in~\cref{eq:hom_CM} and by the secret key rate formula in~\cref{eq:SecretKeyRate}.} However, advantages of the CV scheme over the DV scheme is that CV entanglement swapping is ``easy''. The ``difficult'' parts of the protocol like counting many photons and preparing Fock states concern the purification step only. For key distribution, CV is also straightforward where the end users do quadrature measurements on their final shared state and can distil a secret key after classical error correction and privacy amplification.


\section{Controlled-sum quantum gates}\label{app:CSUM}

\subsection*{Purification using controlled-sum quantum gates}

Purification works by doing QND measurements of total photon number on multiple modes. One can do this using a controlled-SUM (CSUM) quantum gate followed by photon-number measurements.  The CSUM gate is a two-qudit gate which maps the basis states
\begin{figure}[H]
    \centering
    \includegraphics[width=0.3\linewidth]{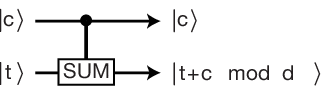},
\end{figure}
\noindent where $\ket{c}$ are basis states of the control and $\ket{t}$ of the target and where $d$ is the dimension of each rail.

Consider that Alice shares $m$ infinite-dimensional maximally-entangled states ($\chi\to1$) with Bob across a lossy channel. Alice and Bob each apply CSUM gates between each of their rails (except the $m$th rail) as control and the final $m$th rail as target. For infinite-dimensional CV states, $d\to\infty$. Alice and Bob destructively measure the photon number of the $m$th rail which is effectively a QND measurement of total photon number of all $m$ rails. For instance, the following circuit locally counts the total number of photons contained in all $m$ modes at Alice and Bob, respectively.
\begin{figure}[H]
    \centering
   \includegraphics[width=0.8\linewidth]{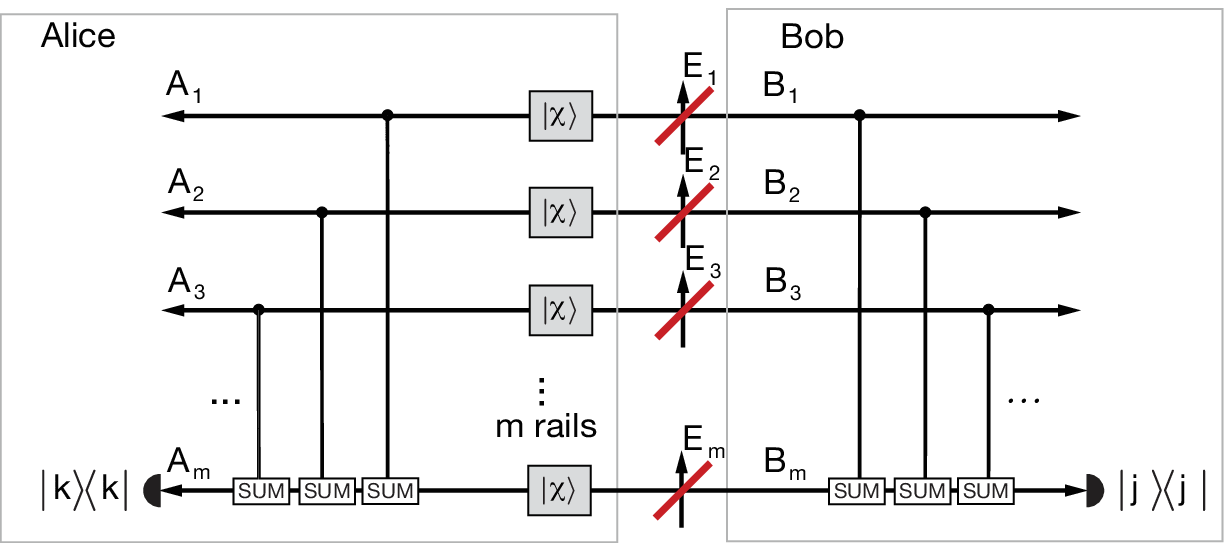},
\end{figure}
\noindent where Alice obtains outcome $k$ and Bob obtains outcomes $j$. This is equivalent to round one of our protocol. Similarly, further gates and measurements can be performed on subsets of rails to iteratively purify entanglement.

Since we provide a linear-optics implementation of our protocol, we naturally have an approximate CSUM gate using linear optics and photon-number measurements \bblack{(without mod $d$)}. Unfortunately, the CSUM gate is distorted, that is, the coefficients on the output state are not ideal. This distortion, however, does not impact purification since we measure the output state right away, and the linear-optics implementation of purification is ideal.

Using linear optics we can do the distorted CSUM:
\begin{figure}[H]
    \centering
    \includegraphics[width=0.4\linewidth]{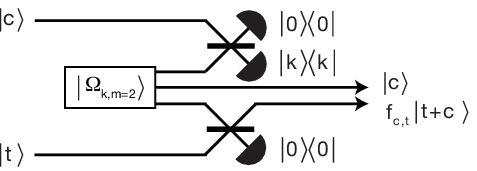},
\end{figure}
where $f_{c,t}$ are in general unwanted factors. 

For $k=1,m=2$, the basis states of the control and target are mapped as follows
\begin{align}
    \ket{0}\ket{0} &\longmapsto \sqrt{2} \ket{0}\ket{0},\\
     \ket{0}\ket{1} &\longmapsto \ket{0}\ket{1},\\
      \ket{1}\ket{0} &\longmapsto \ket{1}\ket{1},\\
       \ket{1}\ket{1} &\longmapsto \ket{1}\ket{2}.\label{eq:linN=1}
\end{align}
There is an unwanted factor $\sqrt{2}$ on the $\ket{0}\ket{0}$ term. For purification this is not an issue since we remove this term via postselection.

Similarly, for $k=2,m=2$, we have
\begin{align}
    \ket{0}\ket{0} &\longmapsto \sqrt{2} \ket{0}\ket{0},\\
     \ket{0}\ket{1} &\longmapsto \ket{0}\ket{1},\\
      \ket{0}\ket{2} &\longmapsto \frac{1}{\sqrt{2}} \ket{0}\ket{2},\\
       \ket{1}\ket{0} &\longmapsto \frac{1}{\sqrt{2}} \ket{1}\ket{1},\\
        \ket{1}\ket{1} &\longmapsto \frac{1}{\sqrt{2}} \ket{1}\ket{2},\\
        \ket{1}\ket{2} &\longmapsto {\frac{\sqrt{3}}{2\sqrt{2}}} \ket{1}\ket{3},\\
        \ket{2}\ket{0} &\longmapsto \frac{1}{\sqrt{2}} \ket{2}\ket{2},\\
         \ket{2}\ket{1} &\longmapsto \frac{\sqrt{3}}{2} \ket{2}\ket{3},\\
         \ket{2}\ket{2} &\longmapsto \frac{\sqrt{3}}{2} \ket{2}\ket{4}.\label{eq:linN=2}
\end{align}

The coefficients of $\ket{\Omega_{k,m=2}}$ are selected so that although the distorted CSUM gate is distorted (which is unavoidable) the purification protocol succeeds perfectly (without distortion).

\subsection*{Single-shot purification via two-mode quantum gates}

The encoding and decoding steps of single-shot purification can be implemented via two-mode quantum gates and photon counting. The code detects loss errors and purifies completely against pure loss in a single shot. For instance, here is the dual-rail ($m=2$) encoding/decoding circuit for all $k$:
\begin{figure}[H]
\centering
\includegraphics[width=0.23\linewidth]{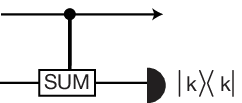},
\end{figure}
\bblack{where ``mod $d$'' plays no role here in the CSUM gate.}

Here is the decoding circuit for $k=1$, $m=3$:
\begin{figure}[H]
\centering
\includegraphics[width=0.5\linewidth]{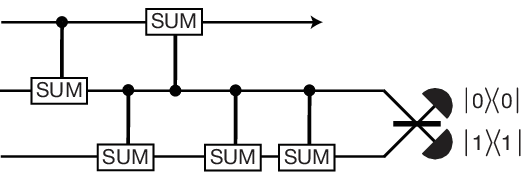}
\end{figure}
\bblack{where the CSUM gates are ``mod 2''.}

\bblack{\section{Implementation using linear optics and photon-number measurements}\label{app:linear_optics}}

\subsection*{Derivation of the circuit}

In this section, we consider how to implement our protocol in an experimentally tractable platform. In this regard, we show that a linear-optical network, coupled with an entangled resource state, implements the required decoding action at Bob's side. We also derive the success probability of this decoding action.

First, consider our ideal single-shot error-detection protocol. Alice prepares the two-mode entangled state $\ket{\psi_{d_{k,m}}}_{A,A1}\propto  \sum_{\mu=0}^{d_{k,m}-1} c_\mu \ket{\mu}_A \ket{\mu}_{A1}$ and encodes the second mode $(A1)$ into our quantum-error-detecting code to send to Bob. Recall that the encoding step is
\begin{align}
    \mathcal{S}_{A1}&= \sum_{\mu=0}^{d_{k,m}-1}  \ket{\mu_{k,m}}_{1,2,3,\dots,m} \bra{\mu}_{A1},
\end{align}
where $\ket{\mu_{k,m}}_{1,2,3,\dots,m}$ are the $m$-rail code words of the code. 
Alice applies it to mode $A1$ of her initial state resulting in
\begin{align}
    (\Id_A \otimes \mathcal{S}_{A1}) \ket{\psi_{d_{k,m}}}_{A,A1}  &\propto  \sum_{\mu=0}^{d_{k,m}-1} c_\mu \ket{\mu}_A \ket{\mu_{k,m}}_{1,2,3,\dots,m}.
\end{align}
After the lossy channels, if Bob detects no loss errors then he successfully decodes, and Alice and Bob share the initial state $\ket{\psi_{d_{k,m}}}_{A,B}$ at the end of the protocol.

Now let us consider the linear-optics circuit, focusing on Bob's side since Alice can encode in a similar way to how Bob decodes, since it is simply the complex conjugate transformation. Alice's encoding is assumed to be done offline and we assume that she has perfectly encoded the quantum information into the $m$-rail code, $(\Id_A \otimes \mathcal{S}_{A1}) \ket{\psi_{d_{k,m}}}_{A,A1}$, and then Bob performs the following action:
\begin{figure}[H]
\centering
\includegraphics[width=0.5\linewidth]{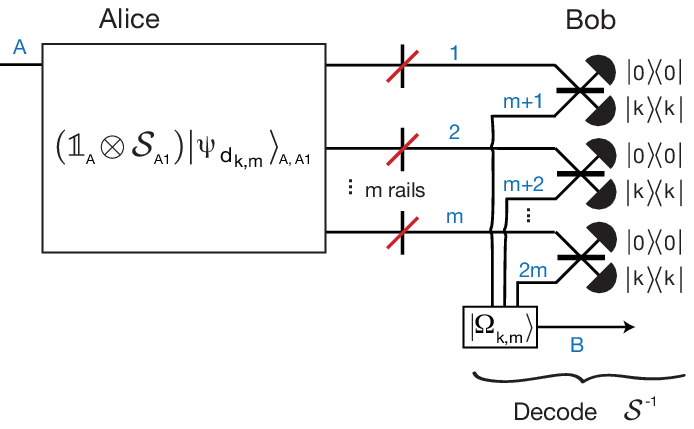}
\end{figure}
This decoding action can be written operationally as follows
\bblack{\begin{align}
    \ket{\psi_{d_{k,m}}} &\propto M_{1,m+1} M_{2,m+2} \cdots M_{m,2m} (\Id_A \otimes \mathcal{S}_{A1}) \ket{\psi_{d_{k,m}}}_{A,A1} \otimes \ket{\Omega_{k,m}}_{m+1,m+2,\dots,2m,B}\\
    &\propto M_{1,m+1} M_{2,m+2} \cdots M_{m,2m}  \sum_{\mu=0}^{d_{k,m}-1} c_\mu \ket{\mu}_A \ket{\mu_{k,m}}_{1,2,\dots,m} \ket{\Omega_{k,m}}_{m+1,m+2,\dots,2m,B}\;,\label{eq:lin_optics_state}
\end{align}
where $M_{u,v}$ are the required measurements on pairs of modes $u$ (sent from Alice through the channel) and $v$ (from Bob's resource state), and $\ket{\Omega_{k,m}}$ is the resource state at Bob's side. The measurements, $M_{u,v}$, consist of mixing modes $u$ and $v$ on a balanced beamsplitter and post-selecting $k$ photons at one output and vacuum at the other
\begin{align}
    M_{u,v} &\equiv \frac{1}{\sqrt{2^k}} \sum_{m=0}^k \sqrt{ k \choose m } \bra{m}_u \bra{k-m}_v.
\end{align}}
This expression can be derived by considering how $|0\rangle|k\rangle$ evolves through a 50:50 beamsplitter, then taking the complex conjugate (see Eq. S2 of~\cite{SUP_guanzon2021ideal}). Recall that the code words in the Fock basis are $\ket{{\mu_{k,m}}}= \ket{n_1^{(\mu)}, n_2^{(\mu)}, n_3^{(\mu)},\cdots,n_m^{(\mu)}}$. Then we define our entanglement resource state as
\begin{align}
    \ket{\Omega_{k,m}}_{m+1,m+2,m+3,\dots,2m,B} \propto \sum_{\mu=0}^{d_{k,m}-1} f_\mu \ket{{\mu_{\widetilde{k,m}}}}_{m+1,m+2,m+3,\dots,2m} \ket{\mu}_{B},
\end{align}
where $\ket{{\mu_{\widetilde{k,m}}}} =  \ket{k{-}n_1^{(\mu)}, k{-}n_2^{(\mu)}, k{-}n_3^{(\mu)},\cdots,k{-}n_m^{(\mu)}}$, and $f_\mu$ is some free scaling factor which we will determine later.

Now, writing~\cref{eq:lin_optics_state} in full we get
\begin{align}
    \ket{\psi_{d_{k,m}}} &\propto \sum_{m_1=0}^k \sqrt{k \choose m_1} \bra{m_1} \bra{k-m_1}\; \sum_{m_2=0}^k \sqrt{k \choose m_2} \bra{m_2} \bra{k-m_2}\; \sum_{m_3=0}^k \sqrt{k \choose m_3} \bra{m_3} \bra{k-m_3}\; \cdots \nonumber \\ \;\;\;\;\; & \;\;\;\;\;\;\sum_{\mu=0}^{d_{k,m}-1}\sum_{\mu'=0}^{d_{k,m}-1} c_\mu f_{\mu'}  \ket{\mu}_A \ket{n_1^{(\mu)}, n_2^{(\mu)}, n_3^{(\mu)},\cdots,n_m^{(\mu)}} \ket{k{-}n_1^{(\mu')}, k{-}n_2^{(\mu')}, k{-}n_3^{(\mu')},\cdots,k{-}n_m^{(\mu')}} \ket{\mu'}_B .\label{eq:lin_optics_state_full}
\end{align}

\bblack{Here, modes labelled $A$ and $B$ are left alone, while the measurement operators act on pairs of modes, one from the channel and one from Bob's resource state.}

The only non-zero terms in~\cref{eq:lin_optics_state_full} are for $m_i=n_i^{(\mu)}$ and $k-m_i=k-n_i^{(\mu')}$, which means that $\mu=\mu'$. Thus, we have shown that these measurement actions and resource entangled state can perform the required decoding action
\begin{align}
    \ket{\psi_{d_{k,m}}}_{A,B} &\propto   \sum_{\mu=0}^{d_{k,m}-1} c_{\mu}f_{\mu}  \sqrt{  {k \choose n_1} {k \choose n_2} \cdots {k \choose n_m}  } \ket{\mu}_A \ket{\mu}_B \\
    &\propto \sum_{\mu=0}^{d_{k,m}-1} c_\mu \ket{\mu}_A \ket{\mu}_B,
\end{align}
in which we get back our original entangled state (but with one arm now at Bob). However, this requires that we set our resource state to have the following factors
\begin{align}
    f_{\mu} &= \left[ {k \choose n_1} {k \choose n_2} \cdots{k \choose n_m} \right]^{-1/2}.
\end{align}

Finally, we note that the success probability of this action, Bob's measurements after loss characterised by transmissivity $\eta$, is given by
\begin{equation}
    P_{k,m}^\text{Bob linear optics} = \frac{\eta^{k}}{2^{m(k-1)}S_{k,m}}\;.\label{eq:lin_optics_P}
\end{equation}
We have an $1/2^{m(k-1)}$ factor from the $m$ beamsplitter measurements $M$, as well as a factor from the normalised resource state
\begin{align}
    S_{k,m} := \left[ \sum_{\mu=0}^{d_{k,m}-1} f_\mu^2 \right].
\end{align}
The final rate of entanglement purification depends on the probability of Bob's side but not Alice's side since her encoding step is assumed to have been prepared offline:
\begin{align}
    E_{k,m}^\text{single shot linear optics} (\eta)&= \frac{1}{m} P_{k,m}^\text{Bob linear optics} E_{k,m}\\
    \\ &= \frac{1}{m} P_{k,m}^\text{Bob linear optics} {\log_2{ {{k+m-1} \choose k}}},
\end{align}
which for $k=1$ is
\begin{align}
    E_{k=1,m}^\text{single shot linear optics} (\eta)&= \frac{\eta }{m^2} {\log_2{ {m \choose k}}}.
\end{align}
In the next section we derive closed expressions for $S_{k,m}$ for $k=1,2,3,4$.

\subsection*{Success probability linear-optics closed expressions}

Let us consider simplifying the following summation
\begin{align*}
    S_{k,m}:=\sum_{i_1+\cdots+i_m=k}^{{k+m-1}\choose k} \frac{1}{ {k \choose i_1} \cdots {k \choose i_m}} = \frac{1}{(k!)^m} \sum_{i_1+\cdots+i_m=k}^{{k+m-1}\choose k} \prod_{j=1}^m i_j!(k-i_j)!
\end{align*}
We note that many of these terms in the summation will be the same value, given a fixed $k$ value. Hence we can get closed expressions by counting the number of terms for each unique value. \bblack{Note that in this following section, the notation $\{n_1,n_2,\cdots,n_m\}$ is shorthand for the set of all distinct permutations of $(n_1,n_2,\cdots,n_m)$.} 

\noindent $\;$

\noindent Given $k=1$:
\begin{align*}
    &\text{for}\; (i_1,\cdots,i_m)\in\{1,0,\cdots,0\} \; \text{has} \; {m \choose 1} \; \text{terms \bblack{for} which} \; \prod_{j=1}^m i_j! (1-i_j)!=1\\
    S_{k=1,m} &= \frac{1}{(1!)^m} \sum_{i_1+\cdots+i_m=1}^{m\choose 1}  \prod_{j=1}^m i_j! (1-i_j)! = \sum_{i_1+\cdots+i_m=1}^{m} 1 = m.
\end{align*}

\noindent Given $k=2$:
\begin{align*}
    &\text{for}\; (i_1,\cdots,i_m)\in\{2,0,\cdots,0\} \; \text{has} \; {m \choose 1} \; \text{terms \bblack{for} which} \; \prod_{j=1}^m i_j! (2-i_j)!=(2!)^m.\\
    &\text{for}\; (i_1,\cdots,i_m)\in\{1,1,\bblack{0,}\cdots,0\} \; \text{has} \; {m \choose 2} \; \text{terms \bblack{for} which} \; \prod_{j=1}^m i_j! (2-i_j)!=(2!)^{m-2}\\
    S_{k=2,m} &= \frac{1}{(2!)^m} \sum_{i_1+\cdots+i_m=2}^{{m+1}\choose 2}  \prod_{j=1}^m i_j! (2-i_j)! = \frac{1}{(2!)^m} \left[ \sum_{j=1}^{m \choose 1} (2!)^m + \sum_{j=1}^{m \choose 2} (2!)^{m-2}  \right]\\
    &= {m \choose 1} +  {m \choose 2} (2!)^{-2} = \frac{1}{8}(7m+m^2). 
\end{align*}

\noindent Given $k=3$:
\begin{align*}
    &\text{for}\; \left(i_{1}, \cdots, i_{m}\right) \in\{3,0, \cdots, 0\} \;\text{has}\; {m \choose 1}\; \text{terms \bblack{for} which}\; \prod_{j=1}^{m} i_{j} !\left(3-i_{j}\right) !=(3 !)^{m}\\
    &\text{for}\; \left(i_{1}, \cdots, i_{m}\right) \in\{2,1,0, \cdots, 0\} \;\text{has}\; {m \choose 1} {{m-1}\choose 1} \; \text{terms \bblack{for} which} \; \prod_{j=1}^{m} i_{j} !\left(3-i_{j}\right) !=(2 !)^{2}(3 !)^{m-2}\\
    &\text{for}\; \left(i_{1}, \cdots, i_{m}\right) \in\{1,1,1,0 \cdots, 0\} \;\text{has}\; {m \choose 3} \;\text{terms \bblack{for} which} \prod_{j=1}^{m} i_{j} !\left(3-i_{j}\right) !=(2 !)^{3}(2 !)^{m-3}\\
    S_{k=3, m}&=\frac{1}{(3 !)^{m}} \sum_{i_{1}+\cdots+i_{m}=3}^{{m+2}\choose 3} \prod_{j=1}^{m} i_{j} ! (3-i_{j})!=\frac{1}{(3 !)^{m}}\left[\sum_{j=1}^{m \choose 1}(3!)^{m}+\sum_{j=1}^{{m \choose 1} {{m-1} \choose 1}} (2 !)^{2}(3 !)^{m-2}+\sum_{j=1}^{ m \choose 3}(2 !)^{3}(3!)^{m-3}\right]\\
    &={m \choose 1}+{m \choose 1} {{m-1} \choose 1} (2!)^{2}(3!)^{-2}+{m \choose 3}(2!)^{3}(3!)^{-3}=\frac{1}{162}(146m+15 m^{2}+m^{3}).
\end{align*}

\noindent Given $k=4$:
\begin{align*}
    &\text{for}\; \left(i_{1}, \cdots, i_{m}\right) \in\{4,0, \cdots, 0\} \;\text{has}\; {m \choose 1}\; \text{terms \bblack{for} which}\; \prod_{j=1}^{m} i_{j} !\left(4-i_{j}\right) !=(4 !)^{m}\\
    &\text{for}\; \left(i_{1}, \cdots, i_{m}\right) \in\{3,1,0, \cdots, 0\} \;\text{has}\; {m \choose 1} {{m-1}\choose 1} \; \text{terms \bblack{for} which} \; \prod_{j=1}^{m} i_{j} !\left(4-i_{j}\right) !=(3 !)^{2}(4 !)^{m-2}\\
    &\text{for}\; \left(i_{1}, \cdots, i_{m}\right) \in\{2,2,0,0 \cdots, 0\} \;\text{has}\; {m \choose 2} \;\text{terms \bblack{for} which} \prod_{j=1}^{m} i_{j} !\left(4-i_{j}\right) !=(2 !)^{4}(4 !)^{m-2}\\
    &\text{for}\; \left(i_{1}, \cdots, i_{m}\right) \in\{2,1,1,0 \cdots, 0\} \;\text{has} \;{m \choose 1} {{m-1} \choose 2} \;\text{terms \bblack{for} which} \prod_{j=1}^{m} i_{j} !\left(4-i_{j}\right) !=(2 !)^{2}(3 !)^{2}(4!)^{m-3}\\
    &\text{for}\; \left(i_{1}, \cdots, i_{m}\right) \in\{1,1,1,1,0 \cdots, 0\} \;\text{has}\; {m \choose 4} \;\text{terms \bblack{for} which} \prod_{j=1}^{m} i_{j} !\left(4-i_{j}\right) !=(3 !)^{4}(4 !)^{m-4}\\
    S_{k=4, m} &={m \choose 1}+{m \choose 1} {{m-1} \choose 1} (3!)^{2}(4!)^{-2}+{m \choose 2}(2!)^{4}(4!)^{-2} + {m \choose 1} {{m-1} \choose 2} (2!)^{2}(3!)^{2}(4!)^{-3}+{m \choose 4}(3!)^{4}(4!)^{-4} \\
    &=\frac{1}{18432}(17198 m+1153 m^{2}+78 m^{3}+3 m^{4}).
\end{align*}

In summary, we have shown that $P_{k,m}^\text{Bob linear optics}=\frac{\eta^{k}}{2^{m(k-1)} S_{k,m}}$ has closed expressions for fixed $k$ as
\begin{align*}
    P_{k=1,m}^\text{Bob linear optics}&=\frac{\eta}{m}, \\
    P_{k=2,m}^\text{Bob linear optics}&=\frac{8 \eta^{2}}{2^{m}\left(7 m+m^{2}\right)}, \\
    P_{k=3,m}^\text{Bob linear optics}&=\frac{162 \eta^{3}}{2^{2 m}\left(146 m+15 m^{2}+m^{3}\right)}, \\
    P_{k=4,m}^\text{Bob linear optics}&=\frac{18432 \eta^{4}}{2^{3 m}\left(17198 m+1153 m^{2}+78 m^{3}+3 m^{4}\right)}.
\end{align*}
In general, it is clear that $S_{k,m}\propto O(m^k)$, however for small $m$ values $S_{k,m}$ will scale more like $m$. This means that the success probability cost of using linear optics will be dominated by the $1/2^{m(k-1)}$ factor.

\subsection*{Experimental imperfections}

Our purification technique is robust to additional losses such as using inefficient detectors. If we consider that Alice has done her encoding offline and has a perfect state preparation, then additional losses at Bob's side, at the resource states and at the detectors, affects the success probability of Bob's side but not the purity of the output state. Therefore, the circuit is robust to both losses in the channel and at the resource state and detectors on Bob's side. All photons must arrive at Bob's detectors in order for the device to succeed, otherwise it fails.

To show this, recall that the output state of the circuit after Bob's measurement is given by~\cref{eq:lin_optics_state_full} and we must have that $m_i=n_i^{(\mu)}$ and $k-m_i=k-n_i^{(\mu')}$, which means that $\mu=\mu'$. Additional losses anywhere on the $m$ rails (i.e., in the channel, at the detectors, or resource state) will decrease the success probability since it is less likely that $m_i=n_i^{(\mu)}$ and $k-m_i=k-n_i^{(\mu')}$ for all $i$, but it will not decrease the purity when the protocol succeeds. Losses on the modes Alice and Bob keep do affect the purity of course.

Interestingly, if the losses are not the same on all the rails, the output state is still pure, but the amount of entanglement is increased or decreased depending on the ratio of the losses on each rail.

While the linear-optics circuit is robust to losses at Bob's detectors and resource states, unfortunately, dark counts, thermal noise, and the use of bucket detectors will in general decrease the purity and fidelity of the output state, however, numerical simulations suggest that practical values still result in good rates. We analyse these effects next.

\bblack{
Here, we analyse the implementation of our single-shot protocol for $k=1$ and $m=2$ using linear optics and photon-number measurements under a thermal-loss channel and with inefficient detectors and dark counts. These imperfections mostly affect the high-loss regime. The linear-optics probability is dominated by the $2^{m(k-1)}$ factor which favours $k=1$. $k=1$ is usually optimal, furthermore, $k=1$ is always optimal in the high-loss regime. So, we only consider $k=1$.
}

\bblack{
Recall that the rate of the $k=1$ protocol using linear optics and photon-number measurements for a pure loss channel and no imperfections is
\begin{align}
    E_{k=1,m}^\text{single shot linear optics} (\eta)&= \frac{\eta }{m^2} {\log_2{ {m \choose k}}}.
\end{align}
}

\begin{figure}
\centering
\includegraphics[width=0.6\linewidth]{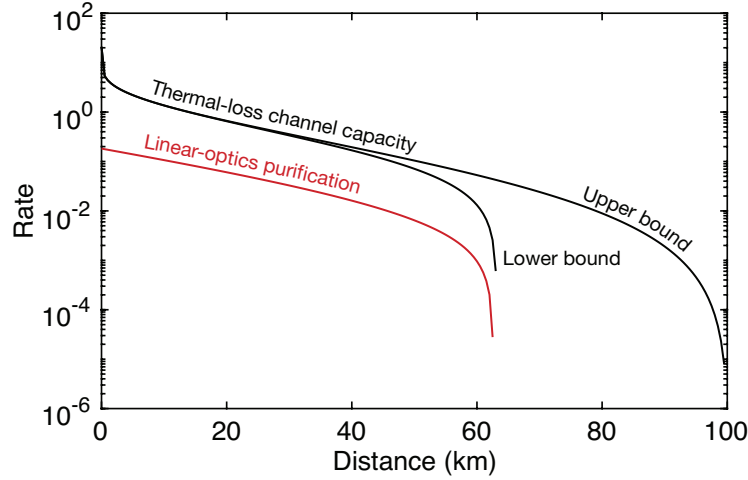}
\caption{\bblack{\textbf{Linear-optics purification with imperfections.} Rate vs. distance for the dual-rail qubit protocol using linear-optics and photon-number measurements. The linear-optics purification rate is given by the RCI. We assume thermal noise in the channel with mean photon number $\bar{n}=0.01$, and imperfections at the detectors with $\eta_\text{efficiency}=0.5$ and dark-count rate $r_\text{dark} = 10^{-6}$. The upper and lower bounds on the channel capacity of the thermal-loss channel are also shown which give the ultimate limits of repeaterless quantum communications~\cite{SUP_Pirandola_2017}.}}\label{fig:dark_counts}
\end{figure}

\bblack{
With inefficient detectors, dark counts, and thermal noise, we perform a numerical calculation of the RCI  and the results are shown in~\cref{fig:dark_counts}. We simulate the thermal-loss channel using the Kraus-operator representation for the channel~\cite{SUP_PhysRevA.84.042311,SUP_PhysRevA.96.062306} with mean photon number $\bar{n}=0.01$. We model dark counts and detector inefficiencies (additional losses) by a thermal-loss channel before each detector at Bob's side with efficiency $\eta_\text{efficiency}=0.5$ and dark-count rate $r_\text{dark}=10^{-6}$. The protocol seems to be tolerant to these imperfections, especially at short distances. Eventually, the thermal noise in the channel overwhelms the protocol. The upper and lower bounds on the channel capacity are also shown~\cite{SUP_Pirandola_2017}.
}

\bblack{\section{Binomial coefficient ratio asymptotics}\label{app:binratioasymp}}

Here, we will prove the asymptotic relation in \cref{eq:binratioasym}, and clearly state the assumptions which contribute to this relation. First, consider the binomial coefficient ratio based on the factorial definition
\begin{align}
   {{k_1+m-1} \choose j_1} { k_1 \choose j_1}^{-1} &= \frac{(k_1+m-1)!}{j_1!(k_1-j_1+m-1)!} \frac{j_1!(k_1-j_1)!}{k_1!} \\ 
   &= \frac{(k_1+m-1)!(k_1-j_1)!}{(k_1-j_1+m-1)!k_1!}. \label{eq:binratioexpansion}
\end{align}
\rred{We will now consider two different methods of proving the required relation.}

\rred{\subsection{Stirling's approximation proof}}
Note that Stirling's approximation $n!\simeq\sqrt{2\pi n} (n/e)^n$ is asymptotically equivalent as $n\to\infty$. Since we are looking at $k_1\to\infty$, we can freely apply Stirling's approximation to these factorials
\begin{align}
    \frac{(k_1+m-1)!(k_1-j_1)!}{(k_1-j_1+m-1)!k_1!} &\simeq \frac{ \sqrt{2\pi(k_1+m-1)} \left(\frac{k_1+m-1}{e}\right)^{k_1+m-1} \sqrt{2\pi(k_1-j_1)} \left(\frac{k_1-j_1}{e}\right)^{k_1-j_1} }{\sqrt{2\pi(k_1-j_1+m-1)} \left(\frac{k_1-j_1+m-1}{e}\right)^{k_1-j_1+m-1} \sqrt{2\pi k_1} \left(\frac{k_1}{e}\right)^{k_1} }\\ 
    &= \frac{ \left(k_1+m-1\right)^{k_1+m-1/2} \left(k_1-j_1\right)^{k_1-j_1+1/2} }{ \left(k_1-j_1+m-1\right)^{k_1-j_1+m-1/2}  k_1^{k_1+1/2} } \\ 
    &= \frac{ k_1^{k_1+m-1/2}\left(1+\frac{m-1}{k_1}\right)^{k_1+m-1/2} \left(k_1-j_1\right)^{k_1-j_1+1/2} }{ (k_1-j_1)^{k_1-j_1+m-1/2}\left(1+\frac{m-1}{k_1-j_1}\right)^{k_1-j_1+m-1/2}  k_1^{k_1+1/2} } \\ 
    &= \left(1-\frac{j_1}{k_1}\right)^{-(m-1)} \left(1+\frac{m-1}{k_1}\right)^{k_1+m-1/2} \left(1+\frac{m-1}{k_1-j_1}\right)^{-(k_1-j_1+m-1/2)}. \label{eq:stirlingeq}
\end{align}
\rred{Recall that we are interested in the large squeezing limit $\chi\to1$ regime, which has two effects. First, Alice will measure a large amount of photons in the first round $k_1\to\infty$. Second, the scaling is done such that there are initially more photons than modes $O(m/f(k_1))\to0$ where $f$ is some polynomial of $k_1$ (for example, $(m-1)/k_1\to0$). Using this, we can show that} the second and third brackets together will go to one. \rred{In particular, note that $\lim_{k_1\to\infty}\left(1+\frac{m-1}{k_1}\right)^{k_1+m-1/2} = e^{m-1}$ and $\lim_{k_1\to\infty}\left(1+\frac{m-1}{k_1-j_1}\right)^{-(k_1-j_1+m-1/2)} = e^{-(m-1)}$. Since both these limits exist, then $\lim_{k_1\to\infty} \left[ \left(1+\frac{m-1}{k_1}\right)^{k_1+m-1/2} \left(1+\frac{m-1}{k_1-j_1}\right)^{-(k_1-j_1+m-1/2)}\right] = e^{m-1}e^{-(m-1)} = 1$. Note with respect to the first brackets in \cref{eq:stirlingeq} that} $j_1$, the amount of photons measured by Bob, is not independent of Alice's measurement\rred{, $k_1$. T}he most likely outcome that Bob will measure is $j_1\approx\eta k_1$, where $\eta$ is the transmissivity of the channel. Hence as the limit is taken, the binomial coefficient ratio will approach 
\begin{align}
    \lim_{k_1\to\infty} \frac{ {{k_1+m-1} \choose j_1} }{{ k_1 \choose j_1}} = \left( 1 - \frac{j_1}{k_1}\right)^{-(m-1)},
\end{align}
as needed to be shown. 

\rred{\subsection{Direct proof}}
\rred{There exists a more direct proof~\cite{SUP_dsingh2022}, starting from \cref{eq:binratioexpansion} we can show 
\begin{align}
    \frac{(k_1+m-1)!}{k_1!} \frac{(k_1-j_1)!}{(k_1-j_1+m-1)!} &= \frac{\prod_{i=1}^{m-1} (k_1 + i)} {\prod_{i=1}^{m-1}(k_1-j_1+i)} = \prod_{i=1}^{m-1} \frac{ 1 + \frac{i}{k_1}} {1-\frac{j_1}{k_1}+\frac{i}{k_1}}. 
\end{align}
Now assuming, in the limit of large squeezing, that $j_1\approx\eta k_1$ (i.e. $j_1/k_1$ is still non-zero in this limit) and that $\lim_{k_1\to\infty}(m-1)/k_1 = 0$ (i.e. in the first round there are more photons than modes), then we can see that
\begin{align}
    \lim_{k_1\to\infty} \frac{ {{k_1+m-1} \choose j_1} }{{ k_1 \choose j_1}} = \lim_{k_1\to\infty} \prod_{i=1}^{m-1} \frac{ 1 + \frac{i}{k_1}} {1-\frac{j_1}{k_1}+\frac{i}{k_1}} = \prod_{i=1}^{m-1} \frac{ 1} {1-\frac{j_1}{k_1}} = \left( 1 - \frac{j_1}{k_1}\right)^{-(m-1)},
\end{align}
as needed to be shown. 
}


%

\end{document}